\documentclass[reprint,superscriptaddress,amsmath,amssymb,aps,prb]{revtex4-2}
\usepackage{xcolor}
\usepackage{graphicx}
\usepackage{dcolumn}
\usepackage{bm}
\usepackage{hyperref}
\usepackage[mathlines]{lineno}
\usepackage[normalem]{ulem}

\begin{document}

\title{Magnetic field induced partially polarized chiral spin liquid in a transition metal dichalcogenide moir\'e system}

\author{Yixuan Huang}
\affiliation{Theoretical Division, Los Alamos National Laboratory, Los Alamos, New Mexico 87545, USA}
\affiliation{Center for Integrated Nanotechnologies, Los Alamos National Laboratory, Los Alamos, New Mexico 87545, USA}

\author{D. N. Sheng}
\affiliation{Department of Physics and Astronomy, California State University, Northridge, California 91330, USA}
 
\author{Jian-Xin Zhu}
\email{jxzhu@lanl.gov}
\affiliation{Theoretical Division, Los Alamos National Laboratory, Los Alamos, New Mexico 87545, USA}
\affiliation{Center for Integrated Nanotechnologies, Los Alamos National Laboratory, Los Alamos, New Mexico 87545, USA}

\date{\today}

\begin{abstract}
As one of the most intriguing states of matter, the chiral spin liquid (CSL) has attracted much scientific interest while its existence and mechanism in crystalline strongly correlated systems remain hotly debated. On the other hand, strong correlation driven emergent phenomena can be realized in twisted transition metal dichalcogenide bilayers with a tremendously tunable large length scale providing a new platform for the emergence of CSLs. We focus on a strongly correlated model relevant to heterobilayer $\textrm{WSe}_{2}/\textrm{MoSe}_{2}$ and investigate the Mott insulating phase at half filling under an out-of-plane magnetic field. Considering both its orbital and spin Zeeman effects we identify three conventionally ordered phases including a $120^{\circ}$ N\'{e}el phase, a stripe phase, and an up-up-down phase. For intermediate fields an emergent quantum spin liquid phase is identified with partial spin polarization. We further characterize its topological nature as the $\nu$ = 1/2 Laughlin CSL through the topological entanglement spectrum and quantized spin pumping under spin flux insertion. In addition, we map out the quantum phase diagram for different twisted angles in an experimentally accessible parameter regime.
\end{abstract}
       
\maketitle

\section{Introduction}
Quantum spin liquids~\cite{balents2010spin,savary2016quantum,zhou2017quantum,broholm2020quantum} (QSLs) are exotic states of matter that feature non-trivial topology~\cite{kalmeyer1987equivalence,wen1991mean,wen2002quantum} and fractionalized excitations~\cite{senthil2002microscopic}. The novel properties of QSLs have attracted numerous studies for several decades which provide potential ways to realize unconventional superconductivity through doping QSLs~\cite{anderson1987resonating,laughlin1988relationship,lee2006doping,fradkin2015colloquium}, and enable topological quantum computation~\cite{kitaev2003fault,nayak2008non}. Furthermore, it has been proposed that the behaviors of many more exotic correlated electronic systems such as the holon Wigner crystal and strange metal can be understood from the perspective of doping  QSLs~\cite{senthil2003fractionalized,patel2016confinement,jiang2017holon}. However, candidate materials of QSLs are very rare partially because of the lack of definite evidence to detect QSLs in experiments.
 
As a promising example, the chiral spin liquid (CSL) carries a quantized chiral edge current that realizes the bosonic version of the fractional quantum Hall effect (FQHE). As a result, CSL may be identified through the quantized thermal Hall conductivity~\cite{kasahara2018majorana, ye2018quantization, bruin2022robustness}. Upon doping, CSL can lead to topological superconductivity~\cite{lee1989} that supports Majorana edge excitations, responsible for edge zero-biased peaks in experiments~\cite{ming2023evidence}. Recently, such novel states have been found numerically on realistic triangular lattice models such as the Hubbard model~\cite{szasz2020chiral,chen2022quantum}, and the extended Heisenberg model~\cite{gong2019chiral} supplemented by three-spin chiral interactions~\cite{wietek2017chiral,gong2017global,zhang20214,kuhlenkamp2022tunable} or four-spin ring exchanges~\cite{cookmeyer2021four}. However, higher-order terms that involve more spins are generally small in perturbation theory and cannot be tuned without considering the interplay among different effective spin interactions. Furthermore, the range of relative strength between nearest neighbor and further neighbor interactions is not fully accessible for most candidate materials that show spin-liquid like behaviors~\cite{shimizu2003spin,kurosaki2005mott,yamashita2008thermodynamic,itou2008quantum,yamashita2009thermal,itou2010instability,zhou2011spin,bordelon2019field}. Thus, the numerical identification for CSLs based on real material parameters remains inconclusive.

Recently, it has been proposed that twisted transition metal dichalcogenide (TMD) bilayers can simulate Hubbard model physics with tunable interacting strengths by different twisted angles~\cite{wu2018hubbard,tang2020,wu2019topological,wang2020correlated,li2021quantum}. In particular, the $\textup{WSe}_{2}/\textup{MoSe}_{2}$ heterobilayer has a single flat band that can be isolated with hole doping, leading to an effective Hubbard model with competing interactions on triangular moir\'e superlattices~\cite{wu2018hubbard}. Even more appealingly, owing to the significantly larger unit cell, an experimentally accessible magnetic field can generate large magnetic flux per unit cell to induce strong effective chiral interactions that are impossible in conventional solid state crystalline systems~\cite{haghshenas2019single,hickey2016haldane,gong2017global,huang2022coexistence,hickey2017emergence}. However, considering the Zeeman term the system energy can be reduced by flipping the spins into a partially polarized state, which makes the non-magnetic CSL, a robust quantum Hall state~\cite{zhang20214}, an excited state. Thus, whether the ground state can be a CSL with a quantized Chern number in a finite magnetization state remains an open question.

Motivated by the possible field induced chiral spin liquids in TMD moir\'e systems, we study the emergent phases of the half-filled Hubbard model in the presence of an applied out-of-plane magnetic field based on the parameters of the $\textup{WSe}_{2}/\textup{MoSe}_{2}$ moir\'e bilayer. Through large-scale density matrix renormalization group (DMRG) simulations~\cite{white1992density,white1993density} and considering both orbital and spin Zeeman effects of the magnetic field, we identify a CSL with partial spin polarization~\cite{kumar2016numerical} at intermediate magnetic fields. In particular, the topological nature of the CSL is identified as the $\nu$ = 1/2 FQHE through spin flux insertion simulations and a topological entanglement spectrum. We also show that the Chern number remains almost quantized in the presence of weak disorders. Applicable to the $\textup{WSe}_{2}/\textup{MoSe}_{2}$ heterobilayer setting, we map out the global quantum phase diagram using experimentally tunable parameters which are twisted angles and magnetic fields. Besides CSL, we identify stripe and up-up-down (UUD) phases with finite magnetization at larger magnetic fields, as well as a $120^{\circ}$ N\'{e}el phase at smaller fields. Our results show that the interplay between orbital and spin effects of the applied magnetic field gives rise to a partially polarized-CSL (PP-CSL), which is stabilized in an experimentally accessible parameter regime. The robust Chern number in finite magnetization and disorder is relevant to the plateau of quantized thermal Hall conductance~\cite{kasahara2018majorana, bruin2022robustness} and can be tested in TMD moir\'e systems.

The rest of the paper is organized as follows: In Sec.~\ref{II}, we introduce the effective spin model and numerical methods used in this work. In Sec.~\ref{III}, we present our main findings in the phase diagram including various magnetic phases and an emergent PP-CSL phase. In Sec.~\ref{IIII}, we further show numerical evidence of the magnetic phases. In Sec.~\ref{IIIII}, we elucidate the nature of the PP-CSL from the entanglement spectrum and the quantized spin pumping via spin flux insertion, which is robust even under a weak disorder. We also discuss the stripelike spin polarization in the CSL phase. In Sec.~\ref{IIIIII}, we give a summary and discussion.

\section{Models and methods}
\label{II}
Considering a heterobilayer with similar lattice constants like $\textup{WSe}_{2}/\textup{MoSe}_{2}$, the moir\'e pattern can be formed with a large moir\'e lattice constant $a_{m}$ and a small twisted angle $\theta_{m}$~\cite{tang2020}. Following Ref.~\cite{wu2018hubbard}, the low-energy bands can be approximated by a single-band Hubbard model with only spin degeneracy, where the Hamiltonian is given as
\begin{eqnarray}
\label{eq_Hubbard}
H_{H} &= \sum\limits_{\left \{ ij \right \} ,\sigma} ( -t_{ij}e^{i\frac{2\pi e}{\hbar}\mathbf{A}\cdot (\mathbf{r}_{i}-\mathbf{r}_{j})}c^{\dagger }_{i\sigma}c_{j\sigma}+\textrm{h.c.}) \\ 
&+U\sum\limits_{i}n_{i\uparrow}n_{i\downarrow} \nonumber +h_{z}\sum\limits_{i}(c^{\dagger }_{i\uparrow}c_{i\uparrow}-c^{\dagger }_{i\downarrow}c_{i\downarrow})\;,
\end{eqnarray}
where we consider the nearest-neighbor (NN) hopping $t_{1}$, next-NN hopping $t_{2}$, and the on-site Coulomb repulsion $U$. An applied out-of-plane magnetic field affects both the orbital as the vector potential $\boldsymbol{\nabla} \times \mathbf{A} = \mathbf{B}$, and the spin as the Zeeman effect $h_{z}=\frac{1}{2}\mu_{B}g_{s}B$, where $\mu_{B}$ is the Bohr magneton and the spin $g$-factor $g_{s}\approx 2$.

\begin{figure}
\centering
\includegraphics[width=1\linewidth]{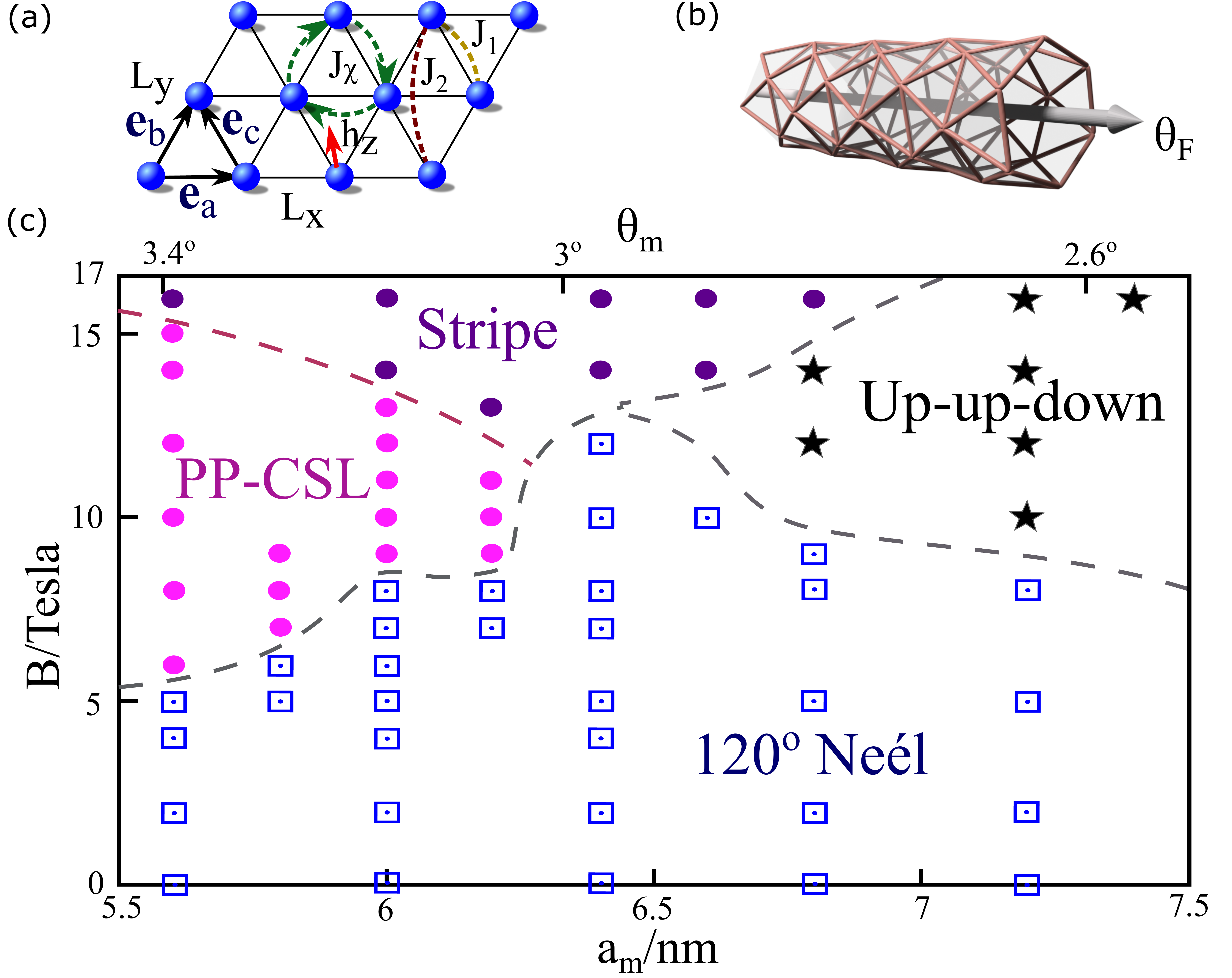}
\caption{(a) Illustration of a triangular lattice with various effective spin interactions. (b) Illustration of flux insertion simulations through the cylinder. (c) Global quantum phase diagram of the moir\'e lattice constant $a_{m}$ and the corresponding twisted angle $\theta _{m}$ vs the magnetic field $B$, obtained on $L_y = 6$ cylinders. We identify  $120^{\circ}$ N\'{e}el, UUD, partially polarized-CSL, and stripe phases. The details of spin values in real space for these phases are presented in SM~\cite{SuppMaterial}.}
\label{Fig_phase_diagram}
\end{figure}

At half filling of the moir\'e bands $U$ is much larger than $t_{1}$~\cite{wu2018hubbard}, thus the charges are localized in the Mott insulating state. The effective spin model can be derived from the perturbation expansion of $t/U$~\cite{macdonald1988t} with effective three-spin chiral interactions~\cite{hickey2016haldane} from the magnetic flux. The overall spin Hamiltonian becomes
\begin{eqnarray}
\label{eq_Spin}
H_{S} &=& J_{1}\sum\limits_{\left\langle ij\right\rangle
}\mathbf{S}_{i}\cdot \mathbf{S}_{j}+J_{2}\sum\limits_{\left\langle \left\langle
ij\right\rangle \right\rangle }\mathbf{S}_{i}\cdot \mathbf{S}_{j} \nonumber \\ 
&+& J_{\chi}\sum\limits_{\{ ijk \} \in \bigtriangleup / \bigtriangledown  }\mathbf{S}_{i}\cdot (\mathbf{S}_{j}\times \mathbf{S}_{k})+2h_{z}\sum\limits_{i}S^{z}_{i}\;.
\end{eqnarray}
Here $\left\langle ij\right\rangle$ and $\left\langle \left\langle ij\right\rangle \right\rangle $ refer to the NN and next-NN site pair, and $\left \{ ijk \right \}$ in the summation refers to three neighboring sites of every unit triangle taken clockwise [Fig.~\ref{Fig_phase_diagram} (a)]. The strengths in the effective spin couplings are $J_{1}=4t_{1}^{2}/U-28t_{1}^{4}/U^{3}$, $J_{2}=4t_{2}^{2}/U+4t_{1}^{4}/U^{3}$, and $J_{\chi }=24t_{1}^{3}\textup{sin}(e\Phi _{B}/\hbar)/U^{2}$, where $\Phi _{B}=2\pi B\frac{\sqrt{3}}{4}a_{m}^{2}$ is the magnetic flux through each unit triangle with $a_m$ being the lattice constant of the triangular moir\'{e} superlattice. Other chiral interactions originating from ring geometries beyond the minimal triangular can also slightly enhance the chiral effect, which are not included here for simplicity. As a remarkable observation, the enlarged length scale of $a_m$ significantly amplifies the orbital effect of magnetic fields, which in turn enhances chiral spin interactions over that in regular solid state crystalline systems by orders of magnitude.

We adopt the moir\'e lattice constant $a_{m}$ dependence of $t_{1}$, $t_{2}$ and $U$ from Ref.~\cite{wu2018hubbard}. As such, the coupling parameters in Eq.~(\ref{eq_Spin}) are ultimately determined by $a_{m}$ and $B$; see more details in the Supplemental Material (SM)~\cite{SuppMaterial}. We explore different phases by tuning $a_{m}$ between 5.5 and 7.5 nm, and $B$ from 0 up to 17 T, as explained later.

The numerical results are obtained by both finite and infinite DMRG methods~\cite{schollwock2011density} with $U(1)$ spin symmetry~\cite{itensor,tenpy}. In flux insertion simulations, a flux $\theta _{F}$ is adiabatically inserted through the cylinder, as illustrated in Fig.~\ref{Fig_phase_diagram} (b). The system has an open boundary in the $e_{a}$ or $x$ direction and periodic boundary conditions in the $e_{b}$ or $y$ direction [Fig.~\ref{Fig_phase_diagram} (a)], with the number of sites denoted as $L_x$ and $L_y$, respectively. The total number of sites is $N=L_{x}\times L_{y}$.
We use finite DMRG with relatively large system sizes up to $N=360$ to access more spin sectors and obtain the $B$ dependence of the magnetization. Flux insertion simulations are carried out with infinite DMRG with a smaller unit cell of 144 sites, because flux insertion does not change the magnetization and the spin pumping is more robust with infinite DMRG algorithm.
We mainly focus on the results on $L_{y} = 6$ cylinders and the results show almost no finite size effect in the $x$ direction.
We keep up to bond dimension $M=6000$ to obtain accurate results with numerical truncation error $\epsilon \lesssim 1\times10^{-6}$; see more details in the SM~\cite{SuppMaterial}.

\section{Phase diagram}
\label{III}
The ground states of the $J_{1}$-$J_{2}$-$J_{\chi }$ model have been explored without the Zeeman effect, which results in CSLs with zero magnetization.
Our numerical results are consistent with previous work in the range of parameters studied in Refs.~\cite{gong2017global, wietek2017chiral} with $h_z=0$; see more details in SM~\cite{SuppMaterial}.

\begin{figure}
\centering
\includegraphics[width=1\linewidth]{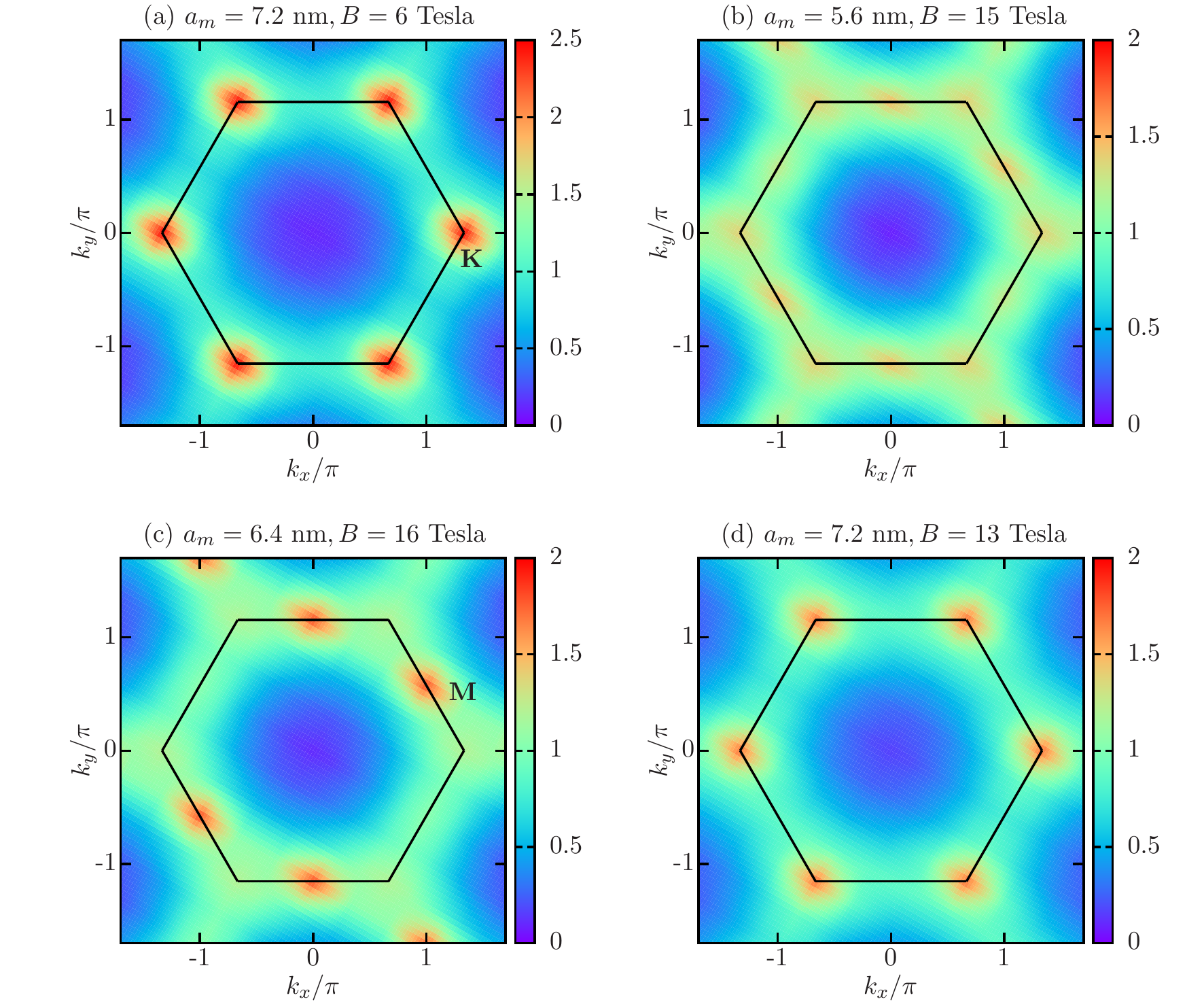}
\caption{Spin structure factor for (a) $120^{\circ}$ N\'{e}el phase, (b) partially polarized-CSL, (c) stripe phase, (d) UUD phase, obtained on $L_{y}=6$ systems. }
\label{Fig_spin_structure}
\end{figure}

\begin{figure}
\centering
\includegraphics[width=1\linewidth]{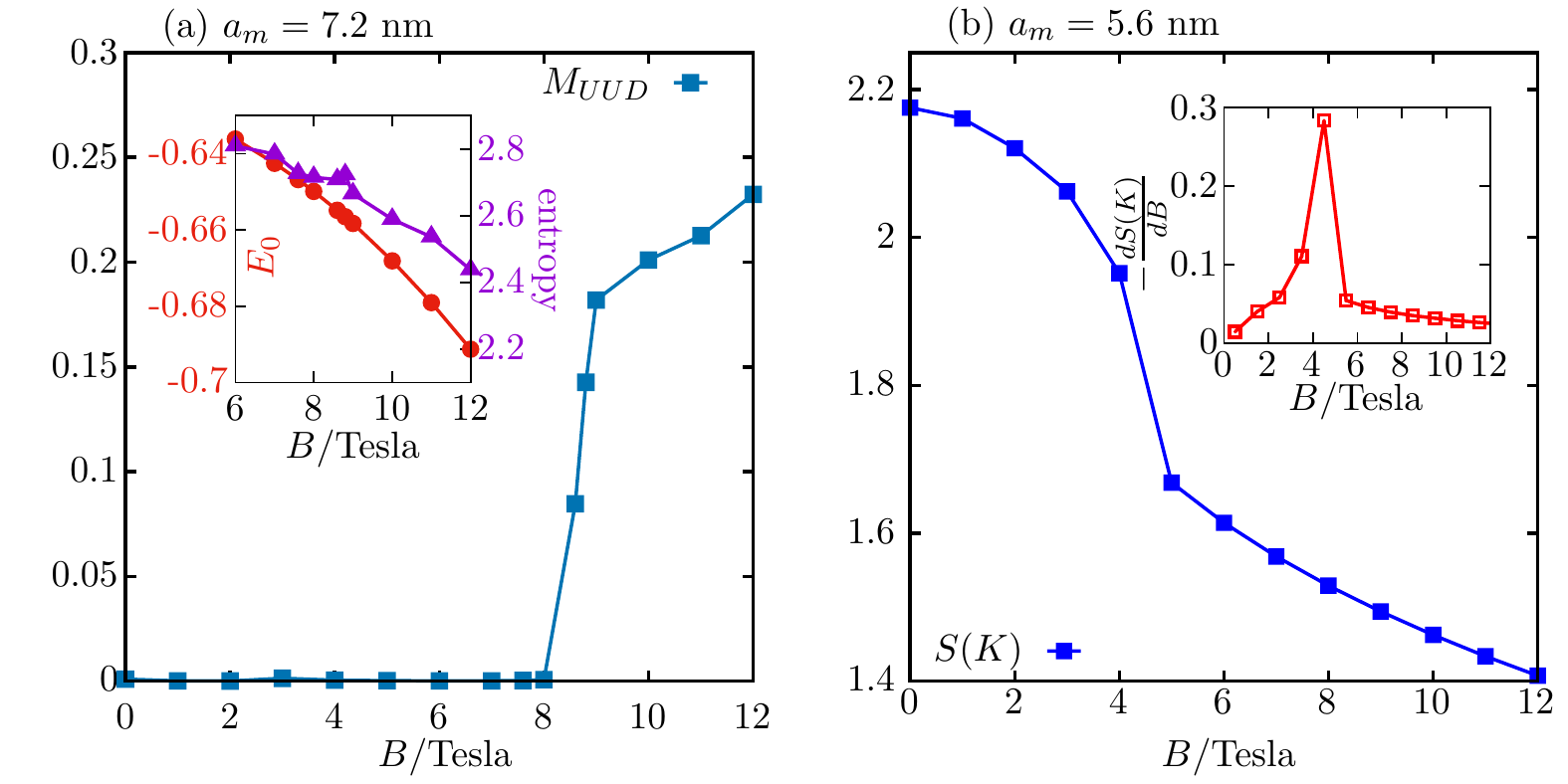}
\caption{(a) Order parameter of the UUD phase for various $B$ at $a_{m}=7.2$ nm with ground state energy per site $E_{0}$ and entanglement entropy in the inset. (b) Structure factor peak for various $B$ at $a_{m}=5.6$ nm. The results are obtained on $L_{y}=6$ systems.}
\label{Fig_order}
\end{figure}

By scanning over the parameter space of Eq.~(\ref{eq_Spin}) we establish a global quantum phase diagram spanned by moir\'e lattice constants and magnetic fields [Fig.~\ref{Fig_phase_diagram} (c)]. We identify three conventional magnetic phases including the $120^{\circ}$ N\'{e}el phase, UUD phase, and stripe phase, as well as a CSL phase. The rich quantum phase diagram results from frustrations and competing interactions, as the classical phase diagram only has N\'{e}el and UUD phases but no stripe phase. The stripe phase may originate purely from the quantum model, where the PP-CSL also exhibit stripelike polarizations; see more details of the classical Monte Carlo results in SM~\cite{SuppMaterial}.
In the limit of $B=0$ T the $120^{\circ}$ N\'{e}el order dominates the whole range of $a_{m}$, where the nearest-neighbor $J_{1}$ takes the leading order, and it smoothly expands to nonzero $B$ regimes. For a smaller value of $a_{m}<5.6$ nm, the 120 N\'{e}el order becomes weak and the ground state might develop into a spin liquid on wider cylinders~\cite{hu2015competing, zhu2015spin} which is beyond the scope of the current work. At larger $B$, we find a stripe phase at small $a_{m}$ and an UUD phase at larger $a_{m}$ before the spins become fully polarized. Both phases exhibit a finite spin polarization in $z$-direction, which increases as $B$ increases. In the intermediate $B$ regime at small $a_{m}$, a CSL with partial polarization emerges. The topological nature of the CSL is identified as the $\nu$ = 1/2 SU(2)$_{1}$ Laughlin state through the flux insertion simulations and the entanglement spectrum. The CSL extends to smaller $a_{m}$, but we are mainly interested in $a_{m}>5.5$ nm, which corresponds to $\theta _{m} \lesssim 3.5^{\circ}$ where the moir\'e band fillings are fully tunable by the electrical gating~\cite{wu2018hubbard}. The partially polarized stripe and UUD phases also extend beyond $B=17$ T in our model. However, the energy scale of Zeeman interactions may interfere with other moir\'e bands, which goes beyond the scope of present work. This is especially true considering the larger $g$-factor for excitons~\cite{forste2020exciton}.

\section{Magnetic orders}
\label{IIII}
The $120^{\circ}$ N\'{e}el phase can be characterized by the static spin structure factor defined as $S(\mathbf{k}) = \frac{1}{N_{m}}\sum_{i,j} (\langle \boldsymbol{S}_{i} \cdot \boldsymbol{S}_{j} \rangle - \langle \boldsymbol{S}_{i} \rangle \langle \boldsymbol{S}_{j} \rangle ) e^{i\mathbf{k}\cdot (\mathbf{r}_{i}-\mathbf{r}_{j})}$, where we sum over $N_{m} = L_{y} \times L_{y}$ middle sites to minimize the boundary effects. $S(\mathbf{k})$ has prominent peaks at the $\mathbf{K}$ points, suggesting the $120^{\circ}$ spin correlations [Fig.~\ref{Fig_spin_structure} (a)]. Tuning finite $B$ into the PP-CSL phase, the peaks disperse along the edge of the Brillouin zone with similar intensities and there is no distinctive peak [Fig.~\ref{Fig_spin_structure}(b)]. By further increasing $B$ into the stripe phase moderate peaks become concentrated at the $\mathbf{M}$ points [Fig.~\ref{Fig_spin_structure}(c)], indicating stripe correlations.

For larger $a_{m}$ in the UUD phase there are mild peaks at the $\mathbf{K}$ points [Fig.~\ref{Fig_spin_structure} (d)]. The magnitude of peaks decreases with increase of $B$ which is consistent with a classical state. In order to determine the UUD order, where spins in the three-sublattice component form $\uparrow \uparrow \downarrow $, we define the order parameter as $M_{UUD}=\frac{1}{N}\sum _{i}\left | \left \langle S^{z}_{i} \right \rangle - \frac{1}{N} \sum _{i}\left \langle S^{z}_{i} \right \rangle \right |$~\cite{li2020partial}. As shown in Fig.~\ref{Fig_order} (a), $M_{UUD}$ increases suddenly around $B=9$ T at $a_{m}=7.2$ nm, indicating a first-order phase transition to the UUD phase. 
The inset of Fig.~\ref{Fig_order}(a) shows the ground state energy as well as the entanglement entropy. The entanglement entropy exhibits a kink near the phase boundary which is consistent with the first-order phase transition. Meanwhile, we find smooth changes in energy and magnetization near the phase boundary; see SM~\cite{SuppMaterial}. The $120^{\circ}$ N\'{e}el order and the UUD order share the same symmetry, therefore we cannot rule out the possibility of phase crossover. 

Figure~\ref{Fig_order} (b) shows the structure factor peak at the $\mathbf{K}$ points for various $B$. Fixing $a_{m}=5.6$ nm, $S(\mathbf{K})$ has a sudden drop at $B=5$ T, indicating the phase transition to CSL. Equivalently it could be seen from the peak in the first order derivative of $S(\mathbf{K})$ in the inset of Fig.~\ref{Fig_order}(b). The phase transition could also be determined by the characteristic entanglement spectrum, which is discussed next.

\begin{figure}
\centering
\includegraphics[width=1\linewidth]{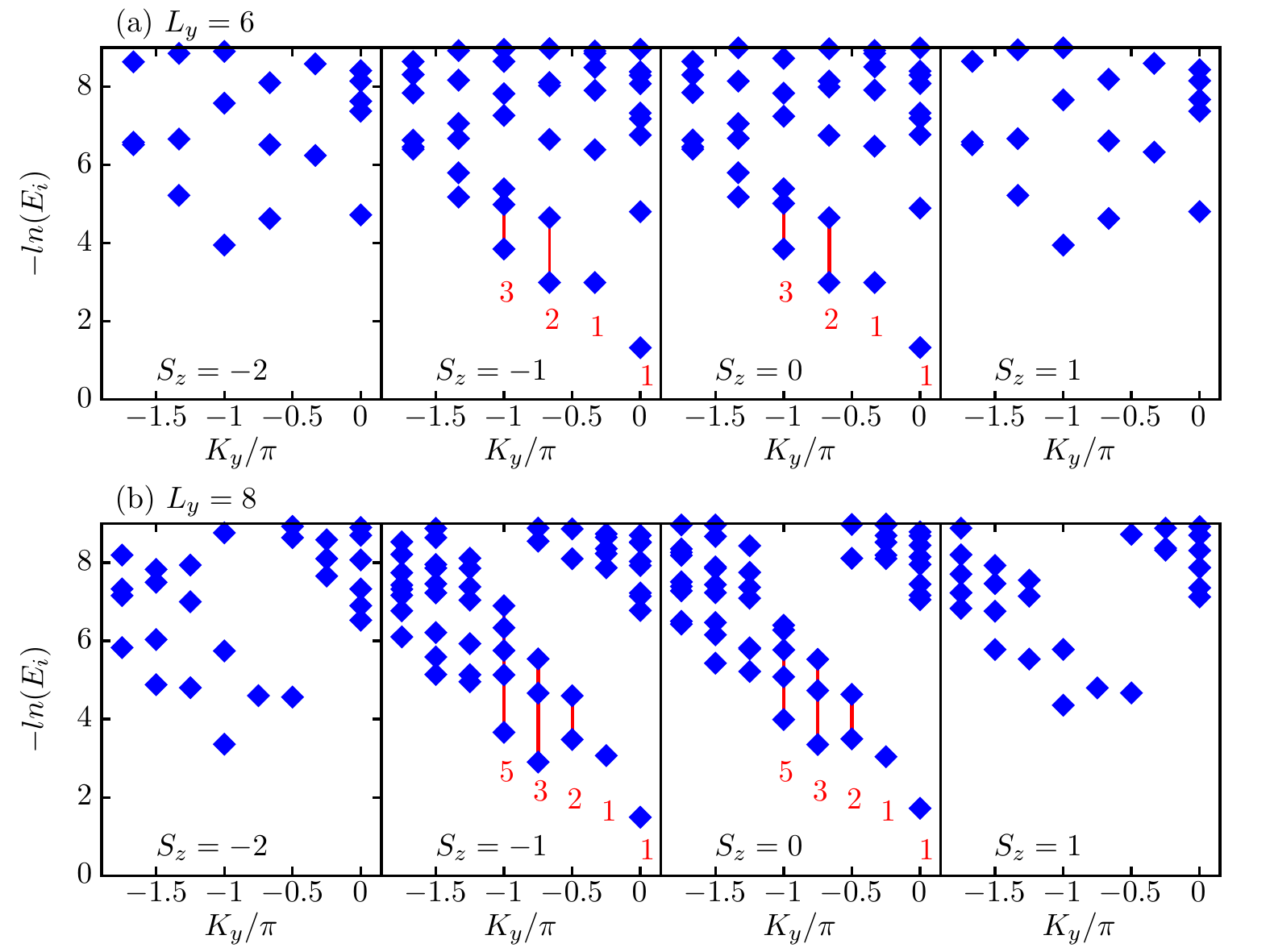}
\caption{Momentum resolved entanglement spectrum at $a_{m} = 5.6$ nm, $B=15$ T in PP-CSL on (a) $L_{y}=6$ and (b) $L_{y}=8$. $K_{y}$ is the conserved discrete momentum in the $y$ direction. Quasi-degenerate eigenvalues are marked by \{1, 1, 2, 3, ...\} in each $S_{z}$ sector.}
\label{Fig_ESK}
\end{figure}

\section{Chiral spin liquids}
\label{IIIII}
The entanglement spectrum extracted from the ground state provides an unbiased way to identify topological states through the one-to-one correspondence to the gapless edge excitation spectrum~\cite{li2008entanglement}. 
In the PP-CSL regime at $a_{m}=5.6$ nm, $B=15$ T, we find that the entanglement spectrum shows quasidegenerate patterns with decreasing momentum on both $L_{y}=6$ and $8$, as shown in Figs.~\ref{Fig_ESK}(a) and \ref{Fig_ESK}(b), respectively. The quasidegenerate eigenvalues form patterns of \{1, 1, 2, 3, ...\} in the lowest two $S_{z}$ sectors, which agrees with the tower of states of the spinon sector~\cite{gong2014,he2014obtaining} by the $SU(2)_{1}$ Wess-Zumino-Witten theory of the $\nu =1/2$ Laughlin state. Higher degenerate levels are not found because of the limited discrete momentum number in finite size lattices. For CSL, the spinon sector is found as the ground state in the presence of magnetic fields, where the vacuum sector has a higher energy.

\begin{figure}
\centering
\includegraphics[width=1\linewidth]{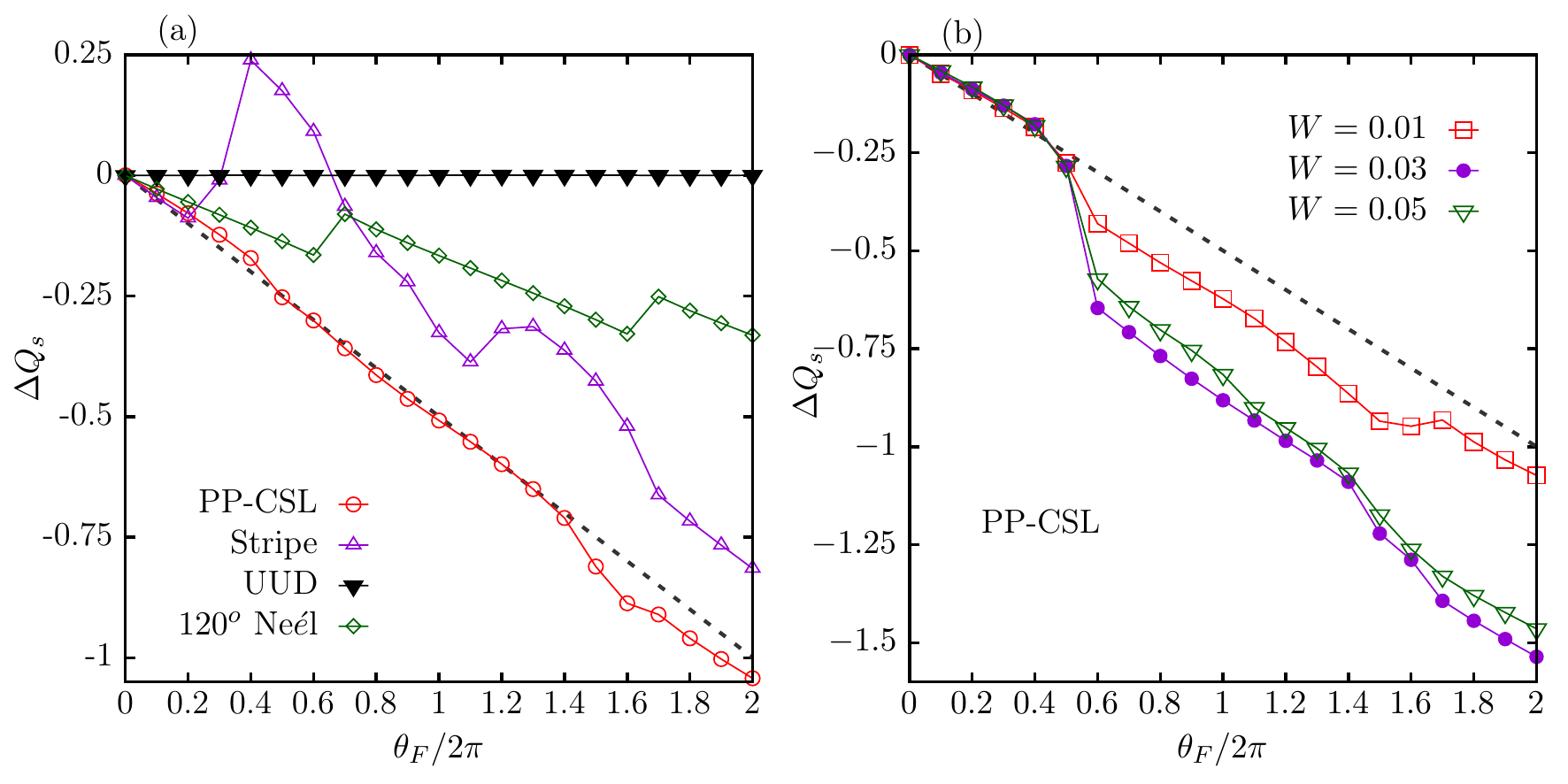}
\caption{(a) Flux insertion simulations in various phases on $L_{y}=6$. For the PP-CSL phase, at 2$\pi$ the spin pumping is almost $\frac{1}{2}$ fractionally quantized. After a $4\pi $ flux the total spin pumping is close to 1. (b) Flux insertion simulations with various disorder in the PP-CSL phase.}
\label{Fig_flux}
\end{figure}

Besides the entanglement spectrum, the spin flux insertion simulation can also identify the topological nature of the CSL~\cite{gong2014,grushin2015characterization}. The spin flux $\theta_{F}$ is adiabatically inserted through the cylinder which adds a phase factor $S_{i}^{+}S_{j}^{-}\rightarrow e^{i\frac{\theta _{F} }{L_{y}}(r^{y}_{j}-r^{y}_{i})}S_{i}^{+}S_{j}^{-}$ to the spin flip terms in the $y$ direction. We measure the pumped spin using the reduced density matrix as $Q_{s} = \sum _{\alpha }\lambda _{\alpha }S^{z}_{\alpha }$ where $\lambda _{\alpha }$ is the eigenvalue and $S^{z}_{\alpha }$ is the corresponding $S^{z}$ of the $\alpha $ eigenstate. Thus, the accumulated spin for different $\theta_{F}$ is $\Delta Q_{s}(\theta _{F})=Q_{s} (\theta _{F}) - Q_{s} (0)$. As shown in Fig.~\ref{Fig_flux} (a), the net spin pumping of $\Delta Q_{s}(\theta _{F} = 2\pi) \approx 0.5$ and $\Delta Q_{s}(\theta _{F} = 4\pi) \approx 1$ which corresponds to a spin Chern number $C=0.5$~\cite{gong2014}, suggesting again the Laughlin type CSL. This is similar to the 1/3 fractional Chern insulator in the Haldane honeycomb lattice model~\cite{grushin2015characterization}. Additionally, we find a disorder-averaged $\Delta Q_{s}(\theta _{F} = 4\pi) \approx 1$ in the presence of weak disorder $W=0.01$ [Fig.~\ref{Fig_flux} (b)] which is defined as $H_{\textrm{disorder}}=\sum_{i}w_{i}S_{i}^{z}$ with $|w_{i}| < W$ being a random effective onsite exchange field; see more details in SM~\cite{SuppMaterial}. At larger disorder $W>0.03$, it deviates from being quantized~\cite{zhu2019disorder}. In the stripe and $120^{\circ}$ N\'{e}el phases the spin pumping becomes not quantized~\cite{huang2022topological} but finite due to a finite non-coplanar chiral order. Additionally, the spin pumping for the classical UUD phase is almost 0. Thus, phase boundaries between the PP-CSL and the stripe phase can be determined by the flux insertion as well as the entanglement spectrum.

Previous studies~\cite{gong2017global} have found a finite spin gap of the non-magnetic CSL without the Zeeman term. We also obtain the spin-1 gap as $\Delta _{s}= E_{0}(S^{total}_{z}=1)-E_{0}(S^{total}_{z}=0)$ in the non-magnetic CSL, which can be used to determine the spin response to the Zeeman field and partial polarization of the state in the thermodynamic limit. The spin gap in the bulk is almost the same for different $L_{x}$, showing little finite size effect. In the thermodynamic limit, if the Zeeman energy of flipping one spin is larger than the finite spin gap, the CSL is partially polarized. We find that the bulk spin gap is smaller in the whole CSL regime, thus the CSL always has partial spin polarization. For example, at $a_{m}=5.6$ nm, $B=15$ T a finite stripelike magnetization in the $z$ direction appears with $\frac{1}{N}\sum_{x,y}(-1)^{x}S_{x,y}^{z}\approx 0.027$; see SM~\cite{SuppMaterial} for more details on the magnetization. The CSL with stripelike polarization and gapless spin excitation is distinctly different from the gapped non-magnetic CSL identified in the previous work~\cite{gong2017global}. Such a stripe has also been found on the honeycomb~\cite{sedrakyan2015spontaneous} and square lattice models~\cite{huang2022coexistence} that coexist with CSL. However, in our results the stripelike spin polarization does not contribute to the transverse spin correlations, and the corresponding spin structure factor shows no distinctive peak in the PP-CSL.

\section{Summary and discussion}
\label{IIIIII}
We have explored magnetic field induced phases in the TMD moir\'e heterobilayer $\textup{WSe}_{2}/\textup{MoSe}_{2}$ by studying a single-band Hubbard model that can be realized for twisted angles $\theta _{m} \lesssim 3.5$~\cite{wu2018hubbard}. Through extensive DMRG simulations on quasi-one-dimensional cylinders, we map out the global quantum phase diagram spanned by twisted angles and magnetic fields. In particular, we have identified a CSL with partial spin polarization for intermediate magnetic fields, and showed its robust Chern number under weak disorder. Furthermore, the partially polarized CSL is surrounded by three magnetic phases including the $120^{\circ}$ N\'{e}el phase, UUD phase, and stripe phase.
The moir\'e superlattices with a tremendously large unit cell reduce the required $B$ to generate strong enough chiral interactions for stabilizing the CSL. The transition from CSL to stripe phase with further increased $B$ can be qualitatively compared to experiments of field induced magnetic transitions of spin-liquid-like phases~\cite{xing2019field}.
For small to intermediate on-site Hubbard repulsion $U$ the electron charge degrees of freedom cannot be neglected. Future studies may explore the field induced phases directly on the Hubbard model, where the CSL may be stabilized for an intermediate $U$ even without $B$~\cite{szasz2020chiral,chen2022quantum}. Besides half filling, other commensurate fillings of moir\'e bands also provide a platform to explore CSL, where Wigner crystals are formed with charges localized in triangular and other types of superlattices~\cite{regan2020mott,huang2021correlated,motruk2023kagome,cai2023signatures,qiu2023interaction}. Finally, we note that our study might also be relevant to a triangular-based spin-$\frac{1}{2}$ metal-organic framework, where the lattice constant can still be much larger than that of typical crystalline solids.

We thank Dr. Fengcheng Wu for helpful discussions. This work was carried out under the auspices of the U.S. Department of Energy (DOE) National Nuclear Security Administration (NNSA) under Contract No. 89233218CNA000001. It was supported by Center for Integrated Nanotechnologies (Y.H.), a DOE BES user facility, in partnership with the LANL Institutional Computing Program for computational resources, and Quantum Science Center (J.-X.Z.), a U.S. DOE Office of Science Quantum Information Science Research Center. D.N.S. was supported by U.S. DOE BES under Grant No. DE-FG02-06ER46305.

\bibliography{CSL_TMD}

%
%

\newcommand{\beginsupplement}{%
        \setcounter{table}{0}
        \renewcommand{\thetable}{S\arabic{table}}%
        \setcounter{figure}{0}
        \renewcommand{\thefigure}{S\arabic{figure}}%
        \setcounter{section}{0}
        \renewcommand{\thesection}{\Roman{section}}%
        \setcounter{equation}{0}
        \renewcommand{\theequation}{S\arabic{equation}}%
        }
\clearpage
\onecolumngrid
\beginsupplement
\setcounter{secnumdepth}{2}

\begin{center}
\Large Supplemental Material for ``Magnetic field induced partially polarized chiral spin liquid in a transition metal dichalcogenide moir\'e system''
\end{center}

\author{Yixuan Huang}
\affiliation{Theoretical Division, Los Alamos National Laboratory, Los Alamos, New Mexico 87545, USA}
\affiliation{Center for Integrated Nanotechnologies, Los Alamos National Laboratory, Los Alamos, New Mexico 87545, USA}

\author{D. N. Sheng}
\affiliation{Department of Physics and Astronomy, California State University, Northridge, California 91330, USA}
 
\author{Jian-Xin Zhu}
\email{jxzhu@lanl.gov}
\affiliation{Theoretical Division, Los Alamos National Laboratory, Los Alamos, New Mexico 87545, USA}
\affiliation{Center for Integrated Nanotechnologies, Los Alamos National Laboratory, Los Alamos, New Mexico 87545, USA}

In the Supplemental Material, we provide more numerical results to support the conclusions we have discussed in the main text. 
In Sec.~\ref{Supp_convergence}, we provide examples of the finite bond dimension scaling of the ground state energy as well as entanglement entropy in the partially polarized-chiral spin lqiuid (PP-CSL) phase and show the numerical convergence of Density Matrix Renormalization Group (DMRG) results. We also show the details of the disorder averaged spin pumping in the presence of a weak disorder.
In Sec.~\ref{Supp_magnetization}, we show the magnetization of different phases in real space.
In Sec.~\ref{Supp_parameters}, we show various spin interactions as a function of the moir\'e lattice constant $a_{m}$ and magnetic field $B$.
In Sec.~\ref{Supp_phase_no_Zeeman}, we show the phase diagram without considering the Zeeman interactions and compare it with our main results.
In Sec.~\ref{Supp_classical_phase}, we implement the classical Monte Carlo methods and obtain the classical phase diagram of the model to compare with our main results.
In Sec.~\ref{Supp_Magnetization}, we show the magnetization evolution with increasing $B$ from the $120^{\circ}$ N\'{e}el phase to Up-up-down (UUD) phase.
In Sec.~\ref{Supp_Chiral}, we show the evolution of the chiral order for various $B$ from the $120^{\circ}$ N\'{e}el phase to PP-CSL phase.

\section{DMRG convergence}
\label{Supp_convergence}

\begin{figure}
\centering
\includegraphics[width=0.95\linewidth]{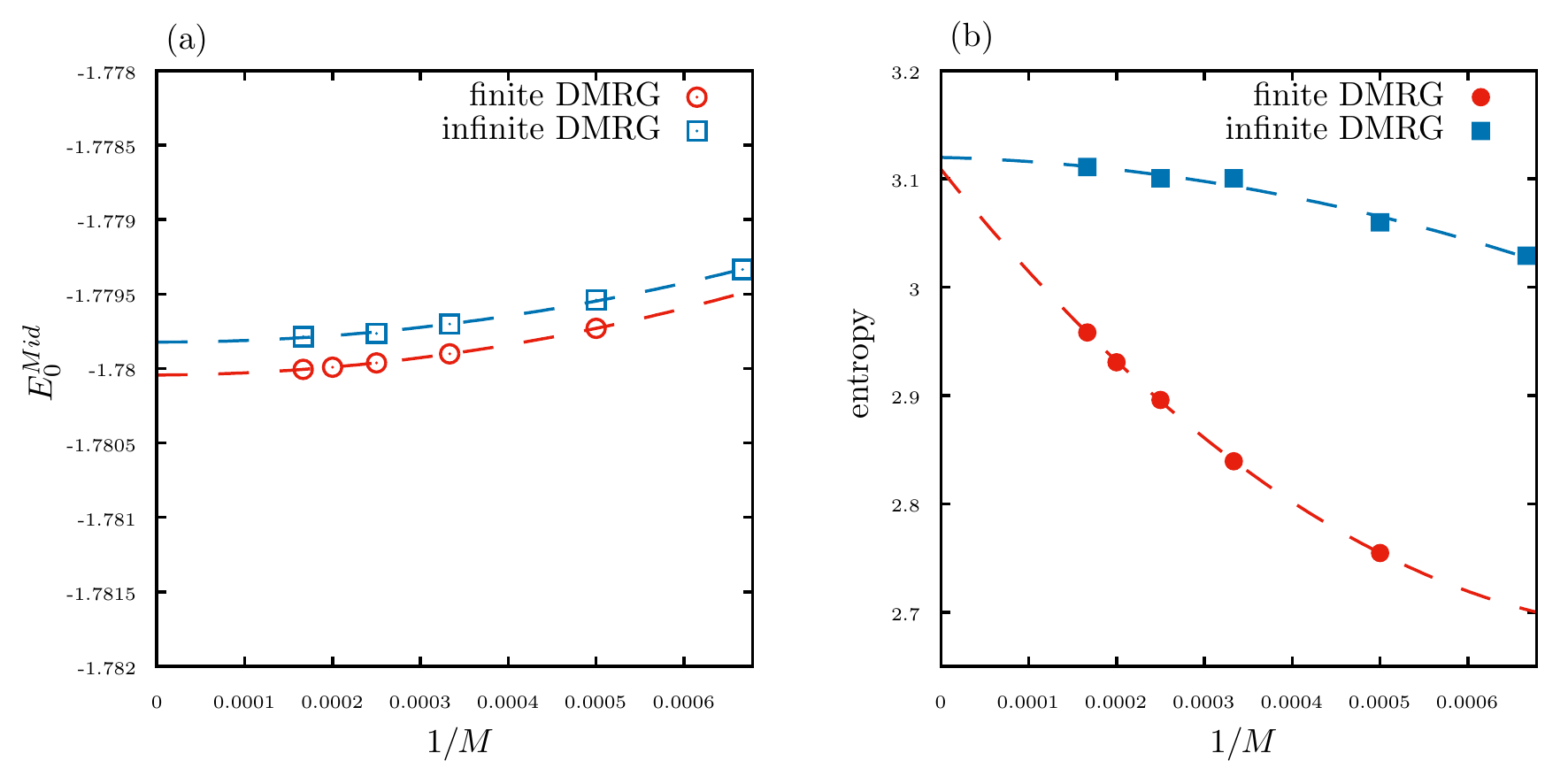}
\caption{Finite bond dimension scaling of the (a) ground state energy per site $E_{0}^{Mid}$ and (b) entanglement entropy in the PP-CSL phase on a $L_{y}=6$ cylinder. The $L_{x}=48$ for finite DMRG results.}
\label{Figs_scaling_M}
\end{figure}

The convergence of DMRG calculation can be examined by the finite bond dimension ($M$) extrapolation of the ground state energy. We show the obtained ground-state energy per site $E_{0}$ versus the inverse DMRG bond dimension ($1/M$) for $L_{y}=6$ from both finite DMRG and infinite DMRG results. For finite DMRG, in order to minimize the boundary effect, we extract the energy using the middle 3/4 (which is $L_x/8 < x \leq  7L_x/8$) of the lattices and calculate the entanglement entropy using the left and right subsystems divided at $L_x/2$.
We keep the bond dimensions up to $M$ = 6000. In Fig.~\ref{Figs_scaling_M} (a) and (b), we show the energies $E_{0}^{Mid}$ and entanglement entropy at $a_{m}=5.6$ nm, $B=15$ T on $L_{y}=6$ with finite and infinite DMRG results, as an example in the PP-CSL phase. The $E_{0}^{Mid}$ and entropy are extrapolated by the second-order polynomials $C(1/M ) = C(0) + a/M + b/M^{2}$, where $C(0)$ is the extrapolated result in the infinite $M$ limit. The energies converge smoothly with bond dimension and the extrapolated energies are very close to the lowest energies we obtain, indicating the good convergence of the results.

\begin{figure}
\centering
\includegraphics[width=0.95\linewidth]{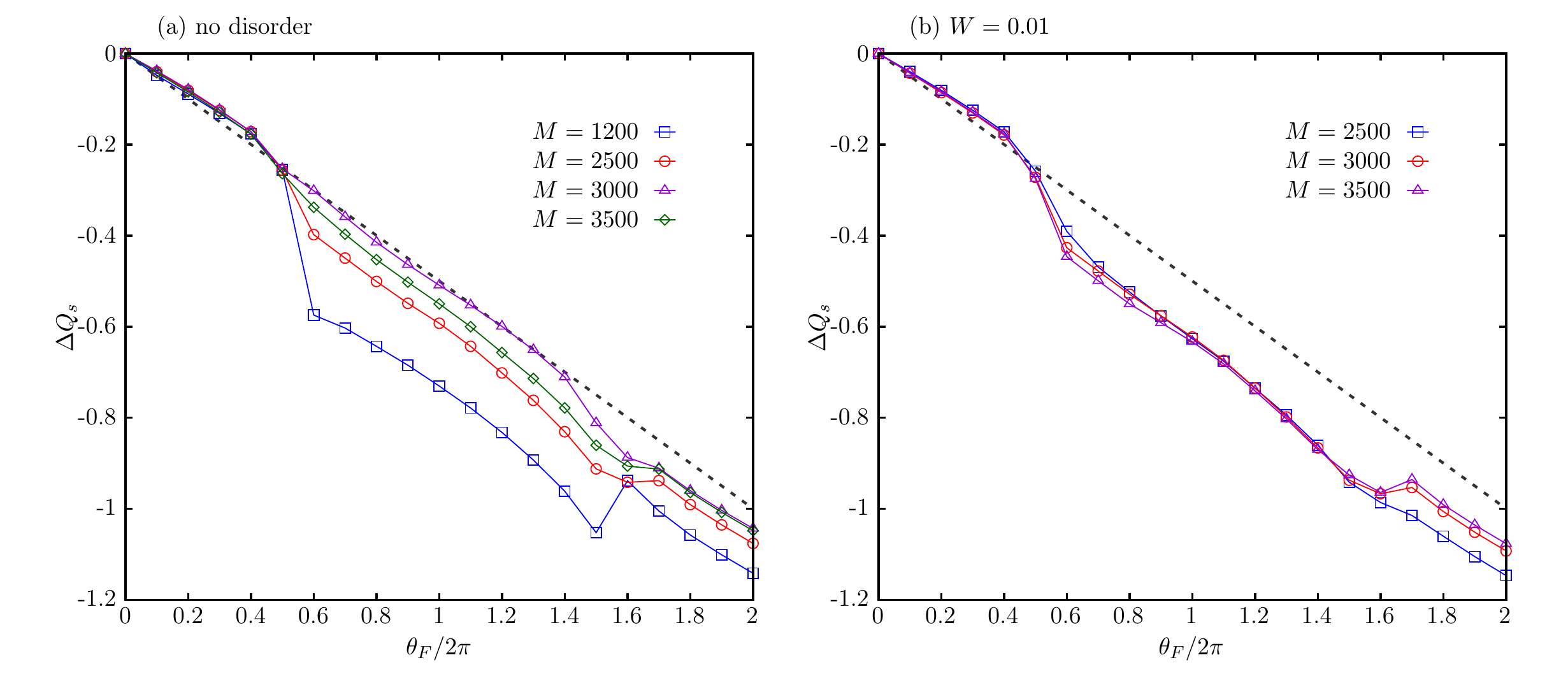}
\caption{The spin flux insertion using various bond dimensions $M$ in the PP-CSL phase at $a_{m}=5.6$ nm, $B=15$ T on a $L_{y}=6$ cylinder with (a) $W=0$ (no disorder) and (b) $W=0.01$.}
\label{Figs_M_flux}
\end{figure}

The spin flux insertion results in the PP-CSL show slight difference for different $M$. As shown in Fig.~\ref{Figs_M_flux}(a), the spin pumping at smaller $M=1200$ have large variations due to a finite magnetization and the pumping is less than 1 at $\theta _{F}=4\pi $. As $M$ increases, the pumping becomes more uniform and the pumped spin is closer to the quantized value of 1.

\begin{figure}
\centering
\includegraphics[width=0.95\linewidth]{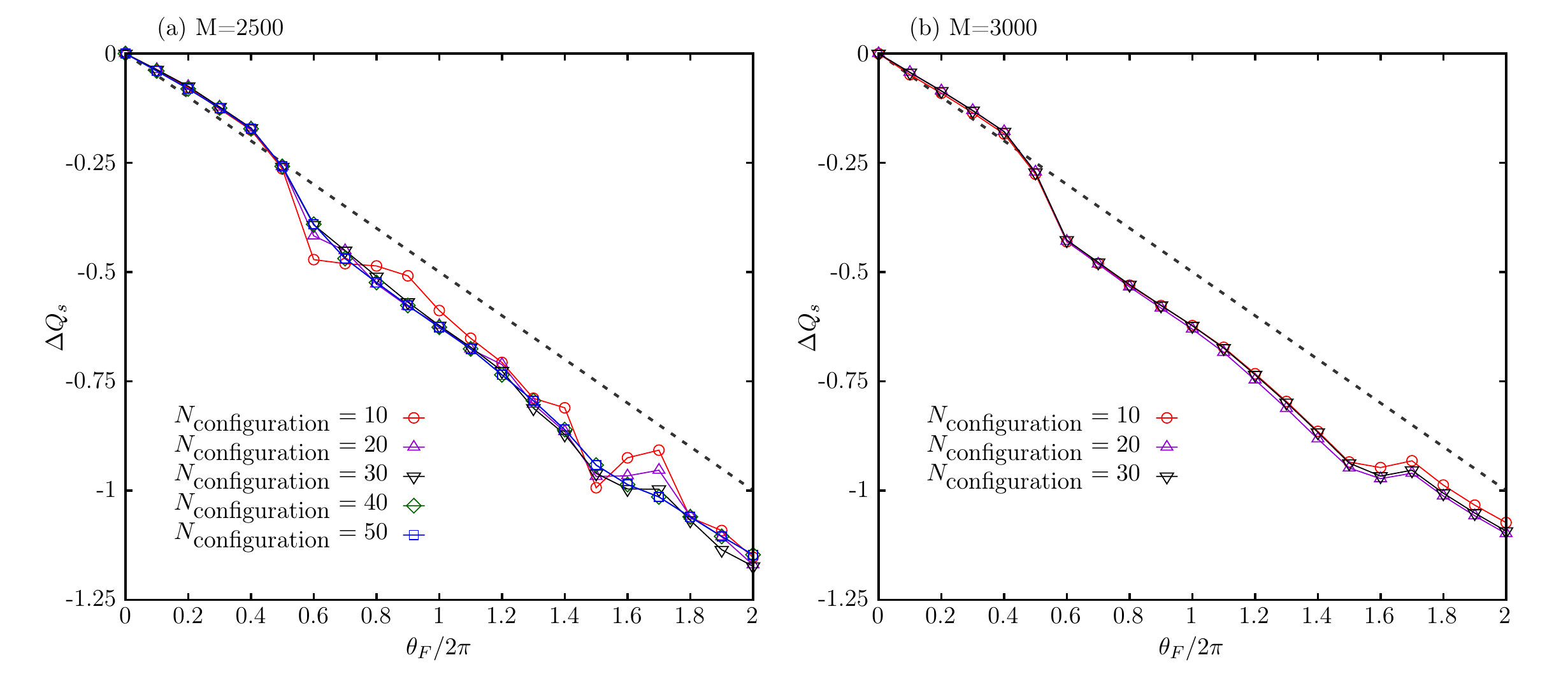}
\caption{The disorder averaged spin flux insertion using various number of random disorder configurations with $W=0.01$ in the PP-CSL phase at $a_{m}=5.6$ nm, $B=15$ T on a $L_{y}=6$ cylinder obtained with (a) $M=2500$ and (b) $M=3000$.}
\label{Figs_flux_trial}
\end{figure}

In the presence of a disorder, we perform an average over random disorder configurations to obtain the averaged spin pumping. The results in the main text are calculated with $M=3000$ and up to 30 disorder configuration averages. To examine the numerical convergence we show the disorder averaged spin pumping $\Delta Q_{s}(\theta _{F})$ for various number of random disorder configurations $N_{\text{configuration}}$ at $W=0.01$ in Fig.~\ref{Figs_flux_trial}. For $M=2500$ in Fig.~\ref{Figs_flux_trial}(a) the $\Delta Q_{s}(\theta _{F}=2\pi)$ remains almost the same for $N_{\text{configuration}} = 10$ to $50$, indicating the converged results for disorder average. Similarly for $M=3000$ in Fig.~\ref{Figs_flux_trial}(b) the disorder averaged $\Delta Q_{s}(\theta _{F})$ is converged with $N_{\text{configuration}}$ up to 30. During the simulations for $W=0.01$, we find that although for a small number of disorder configurations the spin pumping has a sudden drop and becomes much larger than 1, most of the $\Delta Q_{s}(\theta _{F}=2\pi)$ at given random disorder configurations is close to 1. Comparing the results of $M=$2500, 3000, 3500 in Fig.~\ref{Figs_M_flux}(b) we find that using larger bond dimensions can reduce the sudden drop of $\Delta Q_{s}(\theta _{F})$ and lead to the averaged spin pumping of closer to 1. On the other hand, for $W>0.03$ most of the $\Delta Q_{s}(\theta _{F}=2\pi)$ at given random disorder configurations are much larger than 1, which shows distinct difference from the $W=0.01$ results.

\section{Magnetization in real space for different phases}
\label{Supp_magnetization}

\begin{figure}
\centering
\includegraphics[width=0.8\linewidth]{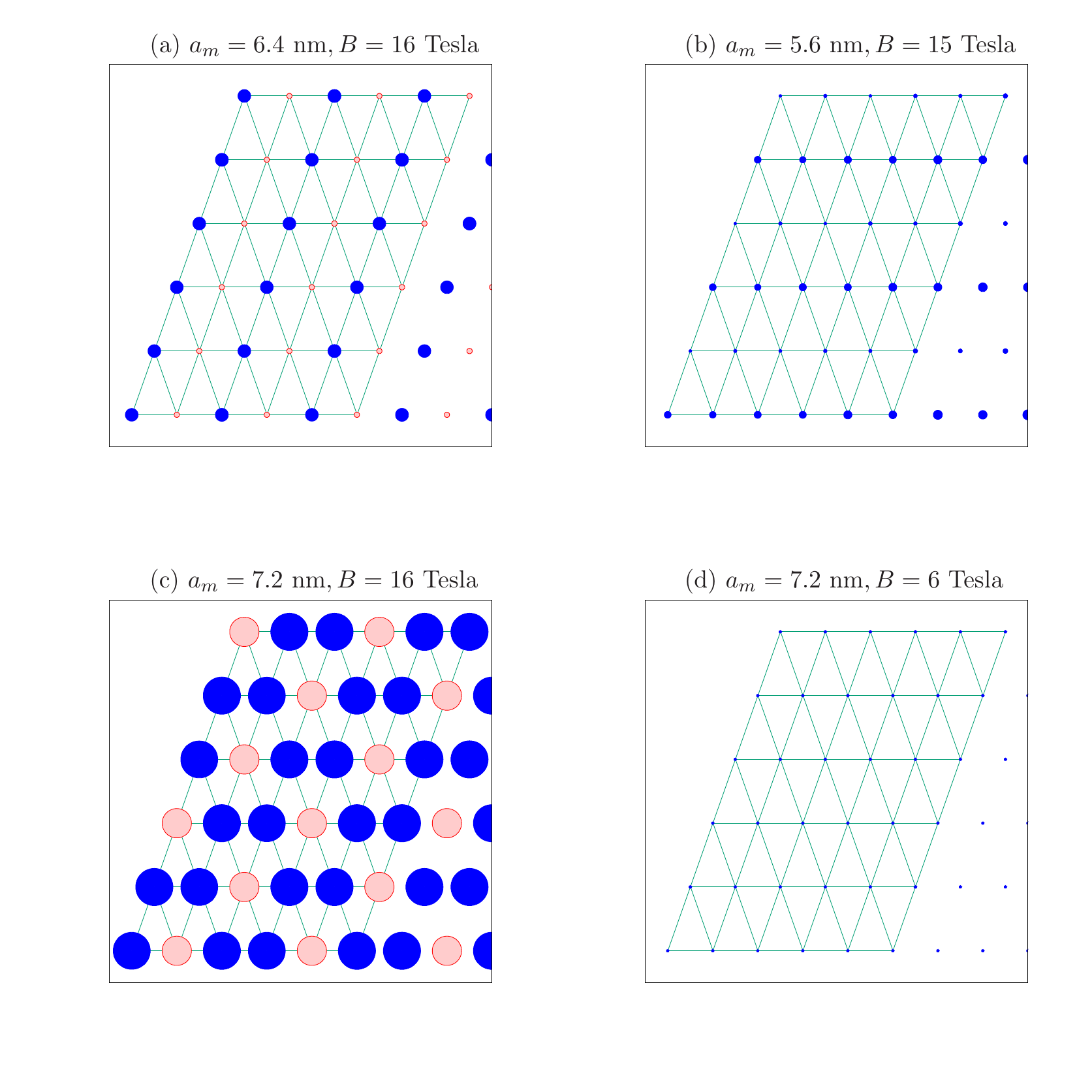}
\caption{Magnetization for (a) the stripe phase, (b) the PP-CSL phase, (c) the UUD phase, and (d) the $120^{\circ}$ N\'{e}el phase. The blue solid circles represent positive spin values, and the red shaded circles represent negative spin values. The radius represents the magnitude where the blue ones in (a), (c), (d) have $\langle S^{z}_{i} \rangle\approx 0.14, 0.41, 0.03$, respectively. The larger blue circles in (b) has $\langle S^{z}_{i} \rangle\approx 0.01$ where the radius is amplified by 10 to see the circle clearer.}
\label{Figs_M}
\end{figure}

We show the magnetization in real space with infinite DMRG for the stripe phase, the PP-CSL phase, the UUD phase, and the $120^{\circ}$ N\'{e}el phase in Fig.~\ref{Figs_M} (a), (b), (c), and (d), respectively. In the stripe phase there are small spin polarization in $z$-direction which is stripe-like as shown in Fig.~\ref{Figs_M}(a), consistent with the stripe spin-spin correlations given in the main text. In the PP-CSL phase [Fig.~\ref{Figs_M}(b)] the stripe-like spin polarization becomes much weaker, and shows a positive spin polarization on every site. We also notice that the stripe changes direction in the PP-CSL phase. In the UUD phase the $z$-component of the spin dominates and the spins in the three-sublattice component form $\uparrow \uparrow \downarrow $, which indicates a classical spin alignment, as shown in Fig.~\ref{Figs_M}(c). For the $120^{\circ}$ N\'{e}el phase at small but finite $B$ we also find a very small spin polarization as shown in Fig.~\ref{Figs_M}(d). The phase is consistent with the $120^{\circ}$ N\'{e}el order at $B=0$ T where the spin-spin correlations are mainly in the $xy$ plane.

\section{$a_m$ and $B$ dependence of spin interaction parameters}
\label{Supp_parameters}

To relate to previous works on the triangular $J_{1}$-$J_{2}$-$J_{\chi }$ spin-1/2 model we show parameters as a function of $a_m$ and $B$. The coupling parameters in the effective spin model are $J_{1}=4t_{1}^{2}/U-28t_{1}^{4}/U^{3}$, $J_{2}=4t_{2}^{2}/U+4t_{1}^{4}/U^{3}$, $J_{\chi }=24t_{1}^{3}\textup{sin}(e\Phi _{B}/\hbar)/U^{2}$, where $\Phi _{B}=2\pi B\frac{\sqrt{3}}{4}a_{m}^{2}$ is the magnetic flux through each unit triangle, and $2h_{z}=\mu_{B}g_{s}B$ where $\mu_{B}$ is the Bohr magneton and the spin g-factor $g_{s}\approx 2$. The $a_m$ and $B$ dependence of $t_1$, $t_2$ and $U$ are obtained from Ref.~\cite{wu2018hubbard}.

\begin{figure}
\centering
\includegraphics[width=1\linewidth]{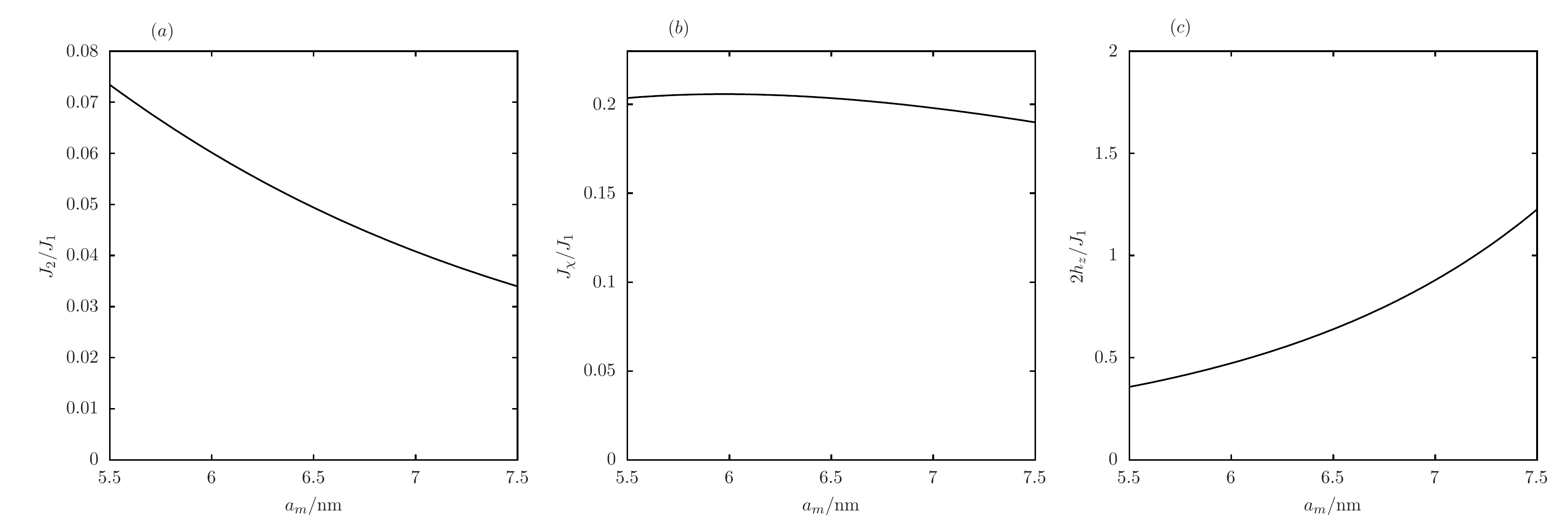}
\caption{Various spin interaction parameters at $B = 10$ T for various $a_m$. (a) shows the $J_{2}/J_{1}$, (b) shows the $J_{\chi }/J_{1}$, and (c) shows the $2h_{z}/J_{1}$.}
\label{Figs_parameter_B}
\end{figure}

Fixing the same $B=10$ T, Fig.~\ref{Figs_parameter_B} (a) shows the $J_{2}/J_{1}$, Fig.~\ref{Figs_parameter_B} (b) shows the $J_{\chi }/J_{1}$, and Fig.~\ref{Figs_parameter_B} (c) shows the $2h_{z}/J_{1}$ for various $a_m$. $J_{2}/J_{1}$ does not depend on $B$, which decreases as $a_m$ increases. $J_{\chi }/J_{1}$ does not have much dependence on $a_m$, mainly because as $a_m$ increases the magnetic flux increases but the interaction strength decreases at the same time. Since the strength $2h_{z}$ does not depend on $a_m$ while  $J_{1}$ decreases as $a_m$ increases, the Zeeman term becomes more dominant at large $a_m$.

\begin{figure}
\centering
\includegraphics[width=1\linewidth]{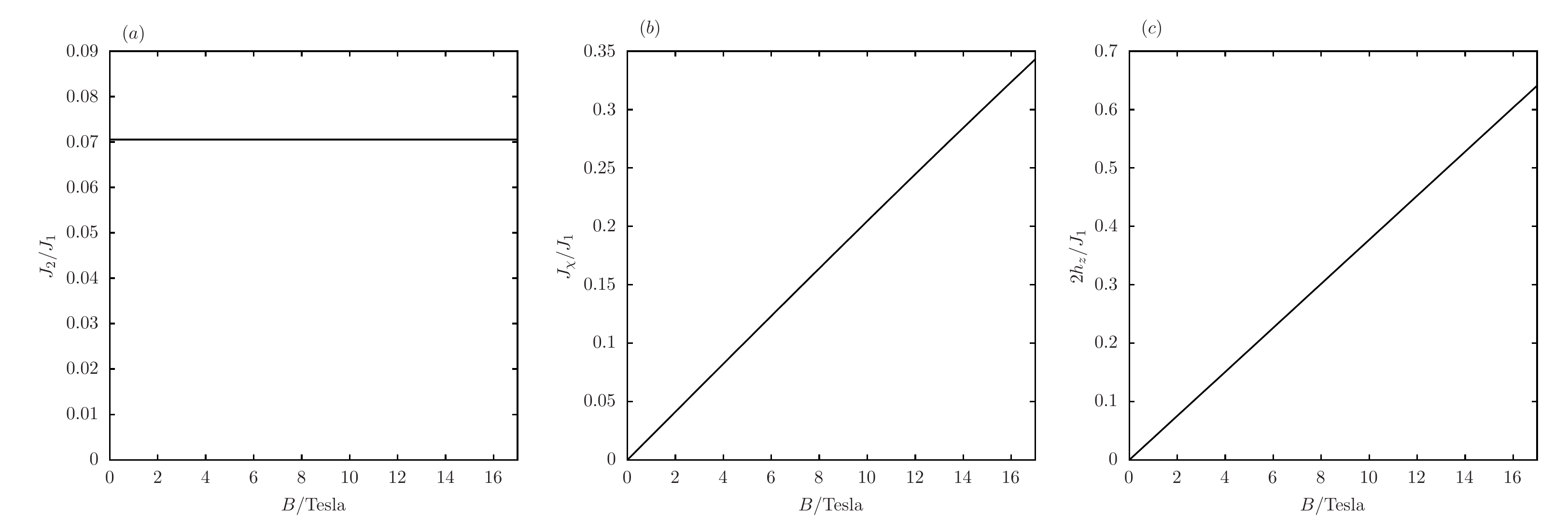}
\caption{Various spin interaction parameters at $a_{m} = 5.6$ T for various $B$. (a) shows the $J_{2}/J_{1}$, (b) shows the $J_{\chi }/J_{1}$, and (c) shows the $2h_{z}/J_{1}$.}
\label{Figs_parameter_am}
\end{figure}

We also show the parameter dependence with fixing $a_{m}=5.6$ nm. Figure~\ref{Figs_parameter_am} shows the same spin interaction parameters as Fig.~\ref{Figs_parameter_B}, where $J_{\chi }/J_{1}$ and $2h_{z}/J_{1}$ increases as $B$ increases, while $J_{2}/J_{1}$ remains the same.

\section{Phase diagram without Zeeman interactions}
\label{Supp_phase_no_Zeeman}

\begin{figure}
\centering
\includegraphics[width=0.6\linewidth]{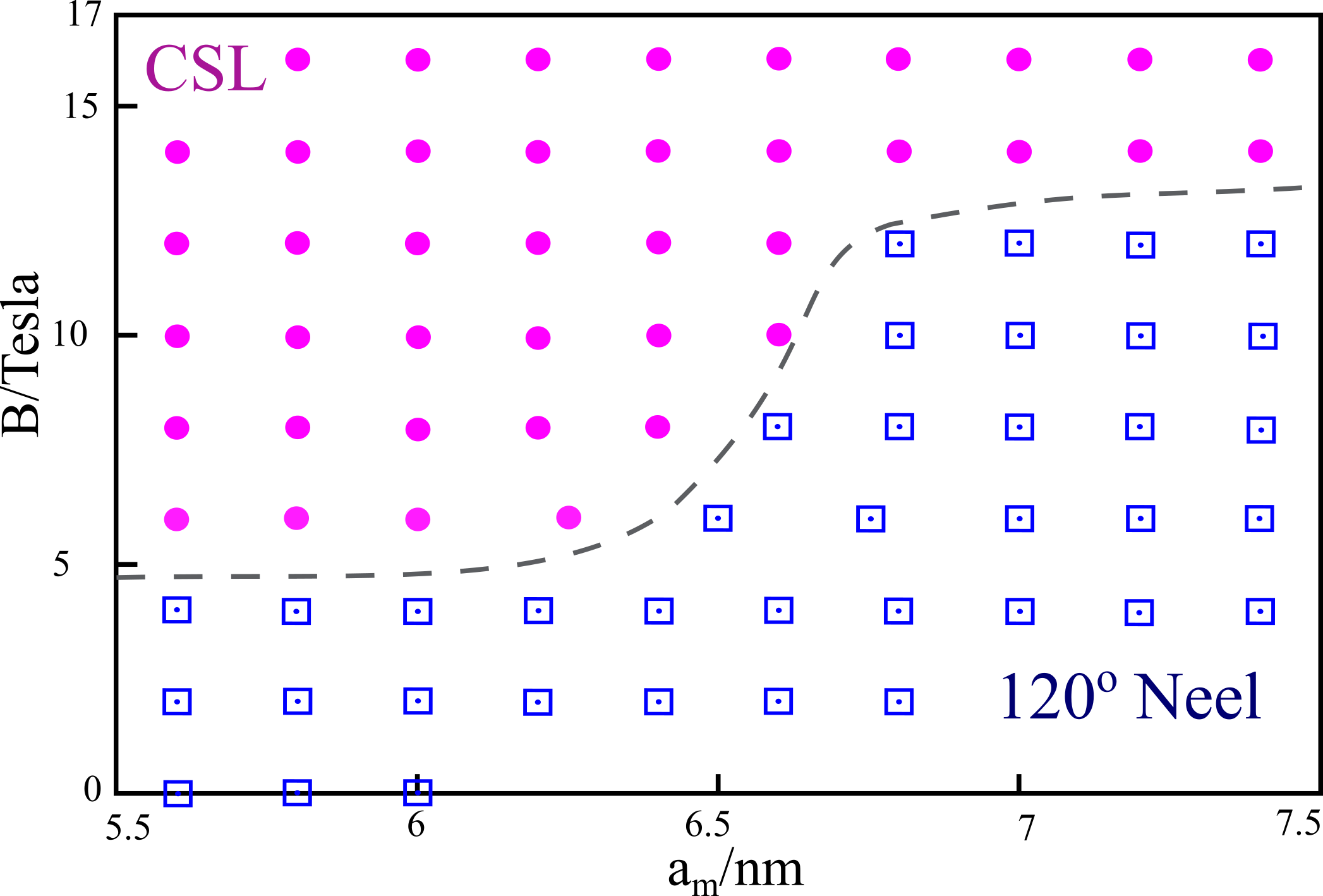}
\caption{The phase diagram without the Zeeman term.}
\label{Figs_phase_no_Zeeman}
\end{figure}

To connect with previous numerical studies of the chiral spin liquids on the triangle lattice with finite chiral interactions, we obtain the phase diagram without considering the Zeeman term ($h_{z}=0$) based on the $L_{y}=6$ system. The resulting $J_{1}$-$J_{2}$-$J_{\chi }$ model has spin SU(2) symmetry and $\sum _{i} S_{i}^{z}=0$ in the ground state with no spin polarization. As shown in Fig.~\ref{Figs_phase_no_Zeeman}, the $120^{\circ}$ N\'{e}el phase is stabilized at small $B$, and CSL at large $B$. For even larger $B$, a tetrahedral order is expected as the chiral interactions become dominant,  leading to a magnetically ordered state~\cite{gong2017global, wietek2017chiral}. Notice that the CSL phase has a larger regime compared to the phase diagram with the Zeeman term, we believe that the spin Zeeman effect of the magnetic field favors the magnetically ordered state while the orbital effect of the magnetic field promotes the CSLs.

\section{Classical phase diagram}
\label{Supp_classical_phase}

To highlight the important quantum fluctuation and correlation effect, we obtain the classical phase diagram of the same model using the classical Monte Carlo methods on the $L_{y} \times L_{x}= 12 \times 12$ system with periodic boundary conditions on both x and y directions. We use the simulated annealing with exponentially decaying temperatures down to 0.01 $meV/k_{B}$ to approximate the ground state. At each temperature we perform $8\times 10^{5}$ sweeps utilizing the \href{https://github.com/fbuessen/SpinMC.jl}{SpinMC package}. 

\begin{figure}
\centering
\includegraphics[width=0.6\linewidth]{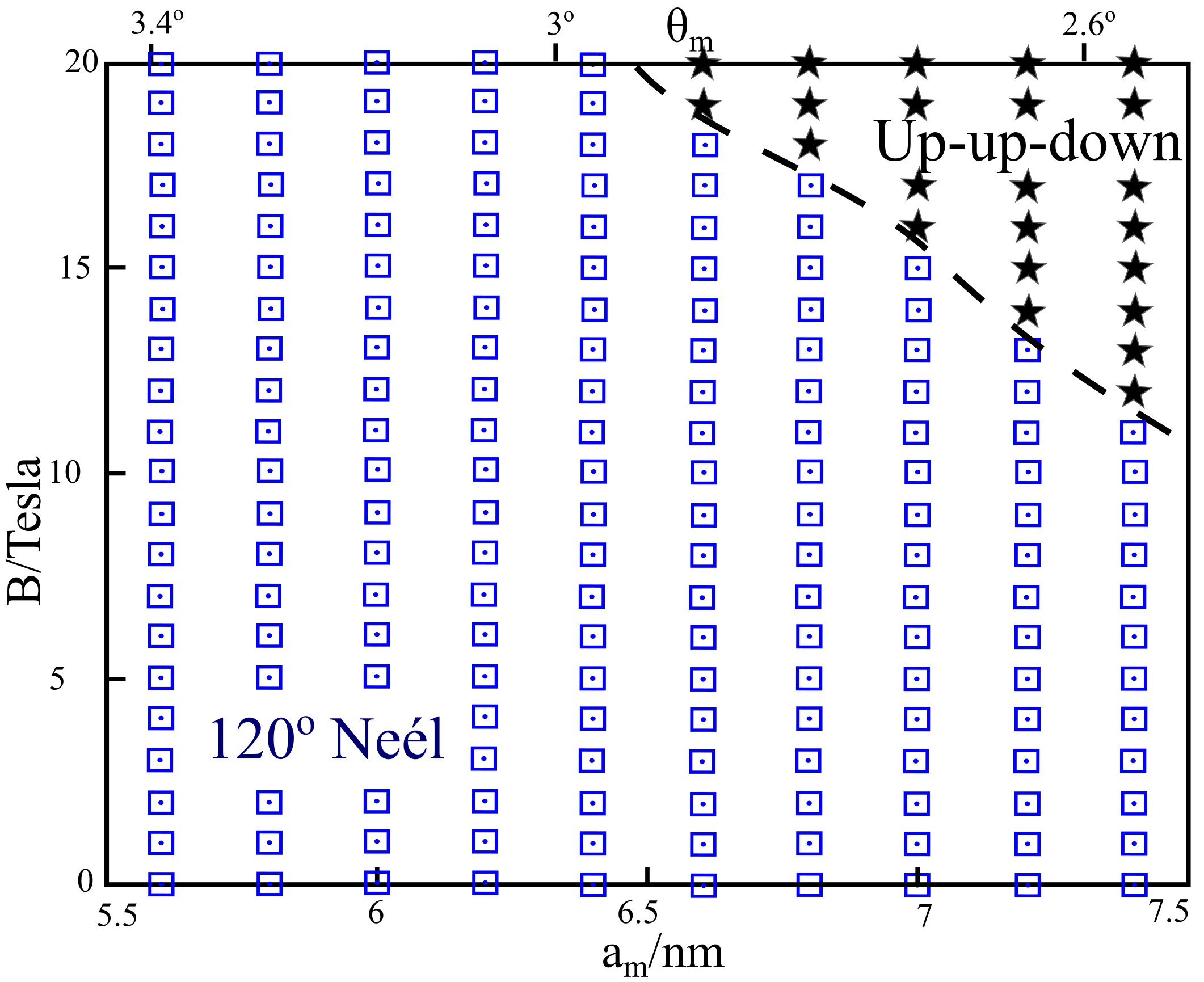}
\caption{The classical phase diagram obtained by the classical Monte Carlo methods. $120^{\circ}$ N\'{e}el phase is identified for small $B$ at all $a_m$, and the Up-up-down phase is identified at large $B$ and $a_m$ similar to the quantum phase diagram. The chiral spin liquid and stripe phase are not identified in the classical phase diagram.}
\label{Figs_classical_phase}
\end{figure}

\begin{figure}
\centering
\includegraphics[width=0.6\linewidth]{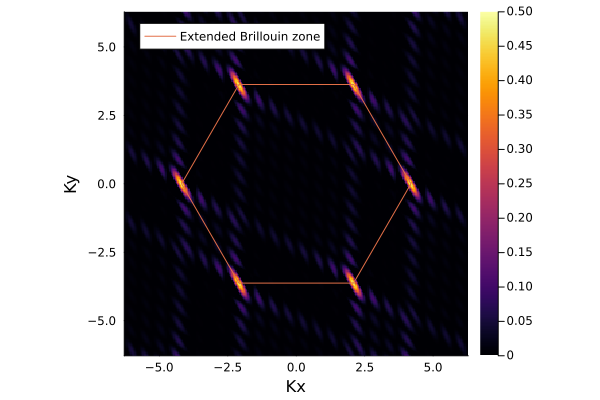}
\caption{The spin structure factor from classical Monte Carlo methods, obtained with $a_{m}=7.2$ nm, $B=0$ T. The red lines indicate the extended Brillouin zone where the peaks are at $K$ points.}
\label{Figs_structure_factor}
\end{figure}

\begin{figure}
\centering
\includegraphics[width=1\linewidth]{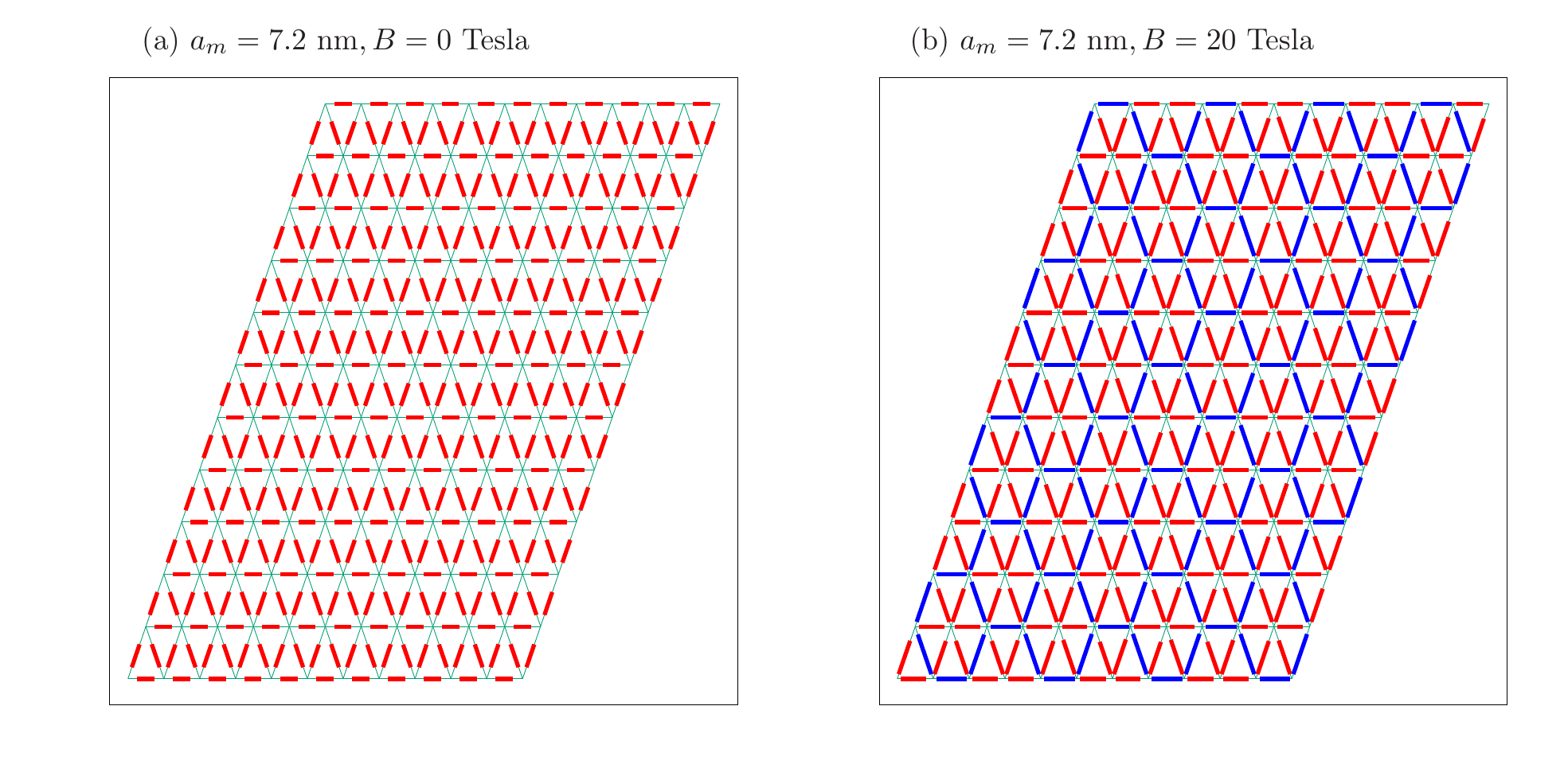}
\caption{The spin-spin correlations between neighboring sites from classical Monte Carlo methods. The $12\times 12$ lattice is drawn with green color. The blue color represent positive spin-spin correlations and the red color represent negative spin-spin correlations between neighboring sites. The length represents the magnitude where the red ones in (a) are around 0.5. The blue ones in (b) are around 0.88.}
\label{Figs_bond}
\end{figure}

As shown in Fig.~\ref{Figs_classical_phase}, the classical phase diagram only has the $120^{\circ}$ N\'{e}el phase stabilized at small $B$, and the Up-up-down phase at large $B$. The chiral spin liquid is a quantum spin liquid that has no classic analog. In addition, the stripe order can only be stabilized considering full quantum fluctuations because the CSL also emerges with the stripe like magnetization. The classical $120^{\circ}$ N\'{e}el phase contains all spins in one plane, which can be rotated freely with the same energy. Using the spin structure factor we find prominent peaks at the $\mathbf{K}$ points suggesting the $120^{\circ}$ spin correlations, as shown in Fig.~\ref{Figs_structure_factor}. We further characterize the classical $120^{\circ}$ N\'{e}el phase by the spin-spin correlations $\mathbf{S}_{i}\cdot \mathbf{S}_{j} = |S_{i}||S_{j}|cos(\theta _{ij})$ between neighboring bonds as shown in Fig.~\ref{Figs_bond} (a). All bonds have negative correlations of around $-0.5$, which corresponds to $\theta _{ij}\approx 120^{\circ}$.

\begin{figure}
\centering
\includegraphics[width=0.6\linewidth]{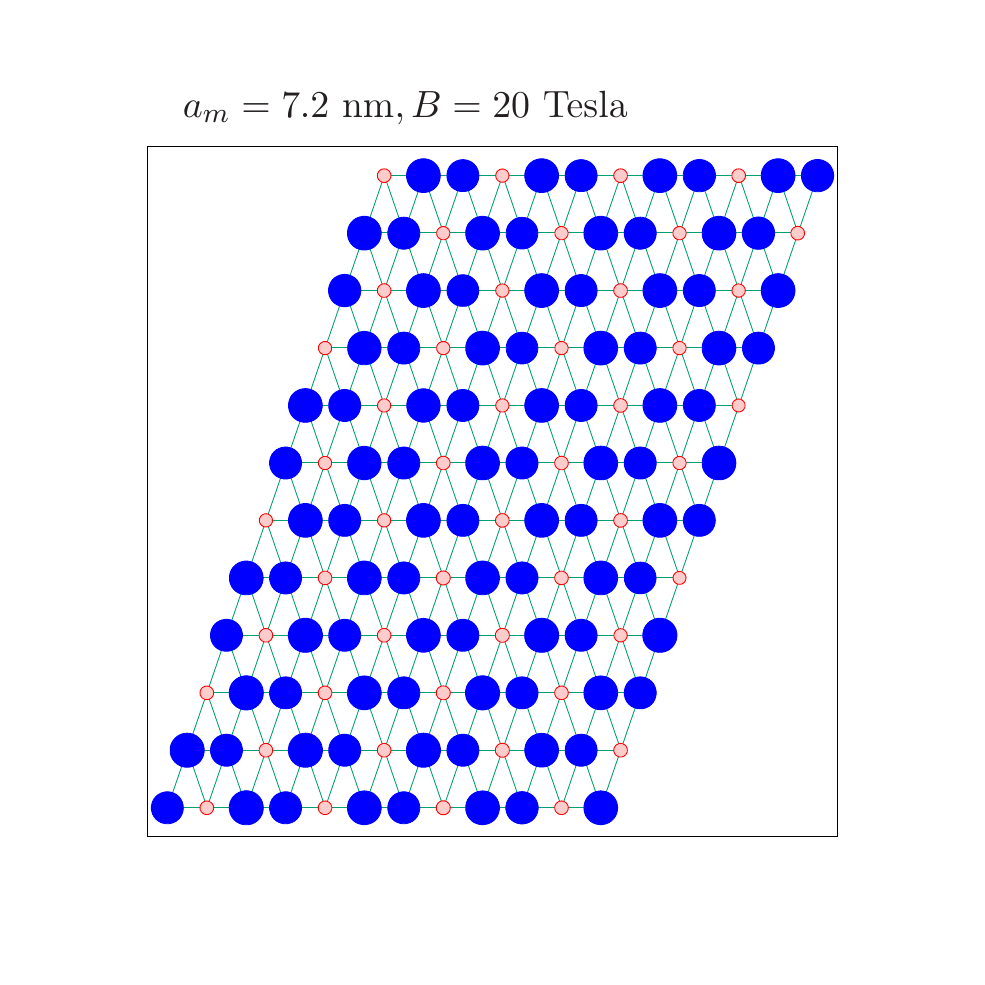}
\caption{The $<S_{i}^{z}>$ from classical Monte Carlo methods, obtained with $a_{m}=7.2$ nm, $B=20$ T. The blue solid circles represent positive spin values, and the red shaded circles represent negative spin values. The radius represents the magnitude where the blue ones are about 0.4.}
\label{Figs_Sz}
\end{figure}

On the other hand, the classical Up-up-down phase at large $B$ has a different pattern of spin correlations as shown in Fig.~\ref{Figs_bond} (b). The correlations along the hexagon are positive indicating the ``Up" spin, and the correlations between them and the hexagonal center are negative, indicating the ``Down" spin at the center. Interestingly, we find that the two ``Up" spins in the classical UUD phase has an angle around $\theta _{ij}\approx 30^{\circ}$. The UUD phase can be further confirmed by the $<S_{i}^{z}>$ value (the magnetic field is also along z direction) as shown in Fig.~\ref{Figs_Sz}, where the spins in the three-sublattice component form $\uparrow \uparrow \downarrow $.

\begin{figure}
\centering
\includegraphics[width=0.9\linewidth]{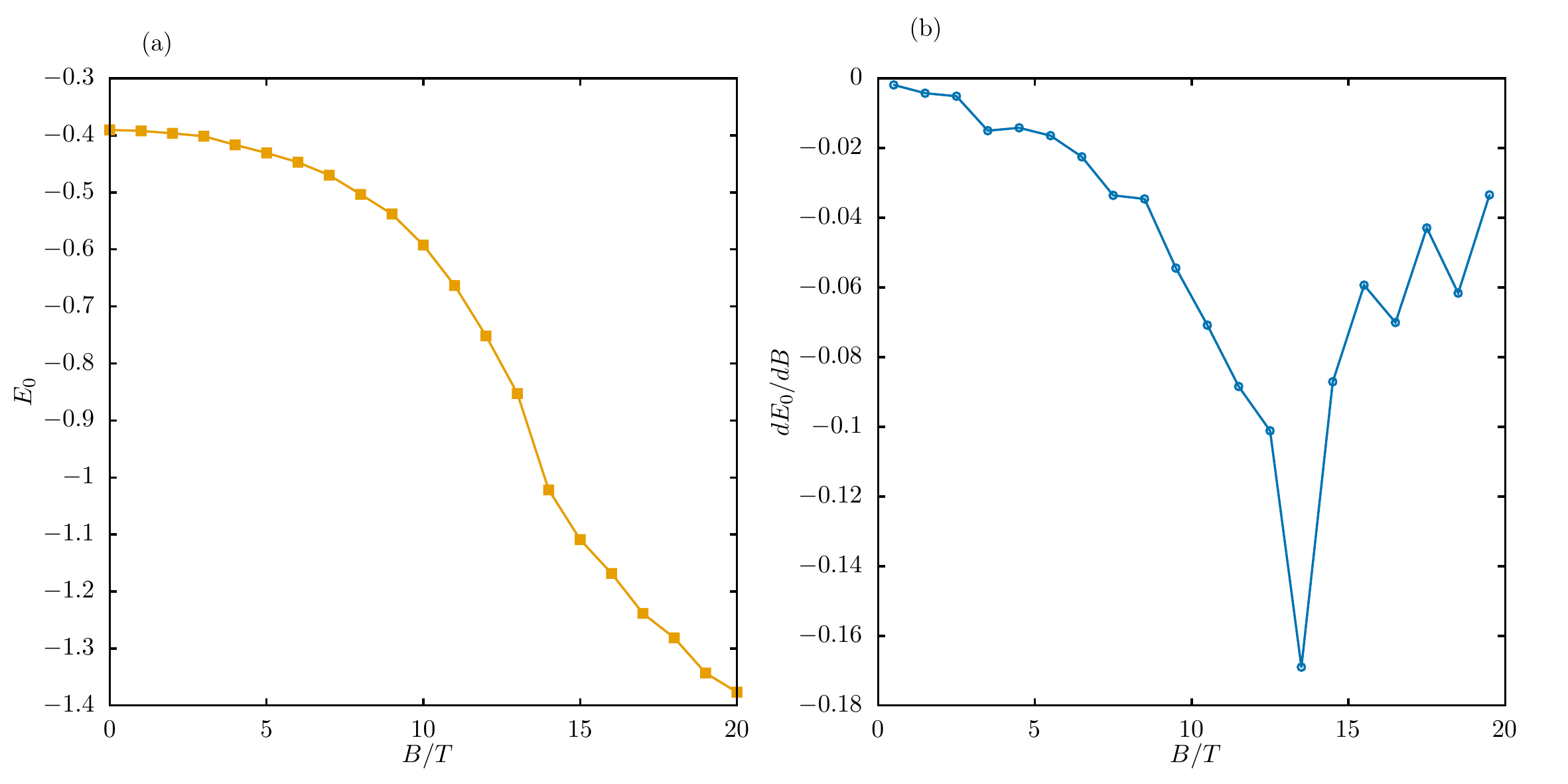}
\caption{(a) The energy per site ($E_{0}$) and (b) the derivative of the energy with respect to $B$ from classical Monte Carlo methods, obtained with $a_{m}=7.2$ nm.}
\label{Figs_M_B}
\end{figure}

To characterize the phase transition between the $120^{\circ}$ N\'{e}el phase and the Up-up-down phase, we obtain the energy per site $E_{0}$ and its first order derivative as a function of $B$. As shown in Fig.~\ref{Figs_M_B} (a), at $a_{m}=7.2$ nm, $E_{0}$ has a sudden drop at $B=13.5$, where its first order derivative shows minimum value,  indicating a phase transition [see Fig.~\ref{Figs_M_B} (b)]. Even though the chiral interactions cannot induce chiral spin liquid in the classical phase diagram, it also plays an important role. Because without the chiral interactions $dE_{0}/dB$ decreases almost linearly over the increase of $B$, which is as expected due to the Zeeman term.

\section{Magnetization from the $120^{\circ}$ N\'{e}el phase to the UUD phase}
\label{Supp_Magnetization}

\begin{figure}
\centering
\includegraphics[width=0.7\linewidth]{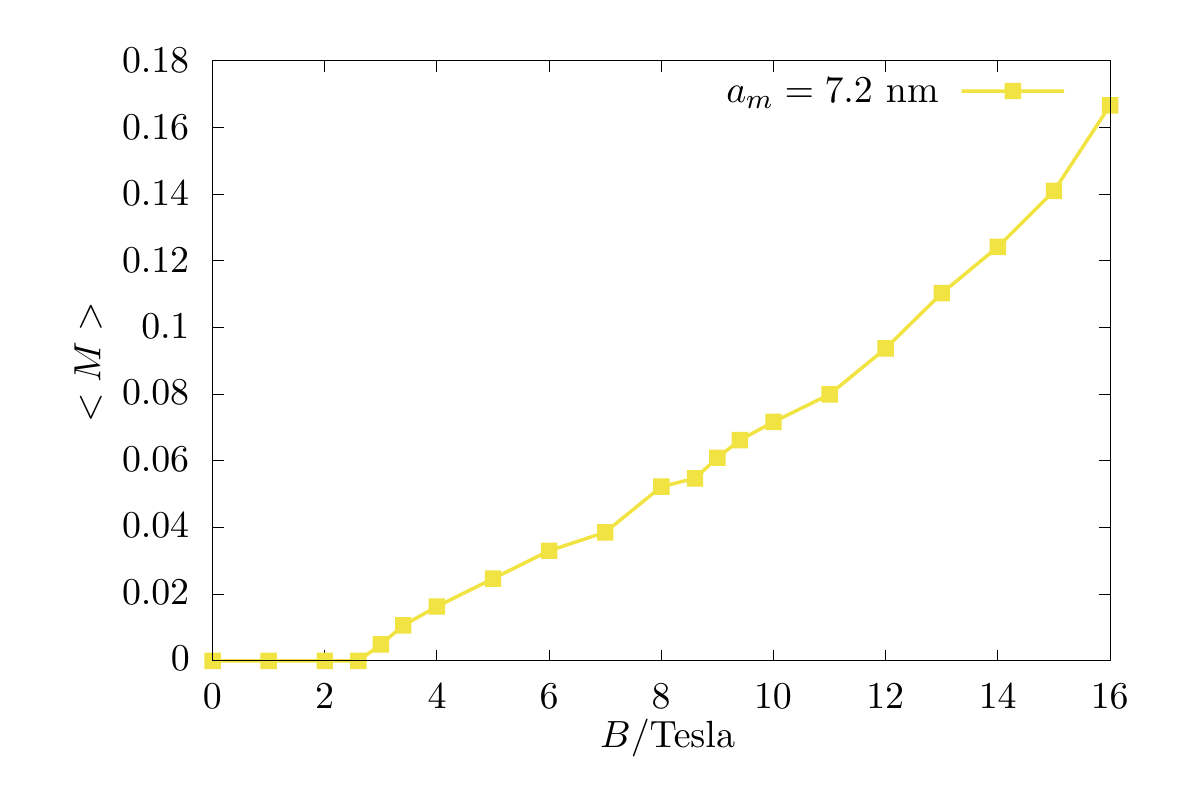}
\caption{ The magnetization $\langle M\rangle$ at $a_{m} = 7.2$ nm for various $B$.}
\label{Figs_MvsB}
\end{figure}

The magnetization is obtained by using the middle half of the lattice to minimize the boundary effect. In order to access more total spin sectors, we use the $N=60\times 6$ lattice with finite DMRG methods. As shown in Fig.~\ref{Figs_MvsB}, the magnetization is 0 in the limit of $B=0$ T and it smoothly increases as $B$ increases from the $120^{\circ}$ N\'{e}el phase to the UUD phase. 

\section{Chiral order from the $120^{\circ}$ N\'{e}el phase to the CSL phase}
\label{Supp_Chiral}

\begin{figure}
\centering
\includegraphics[width=0.7\linewidth]{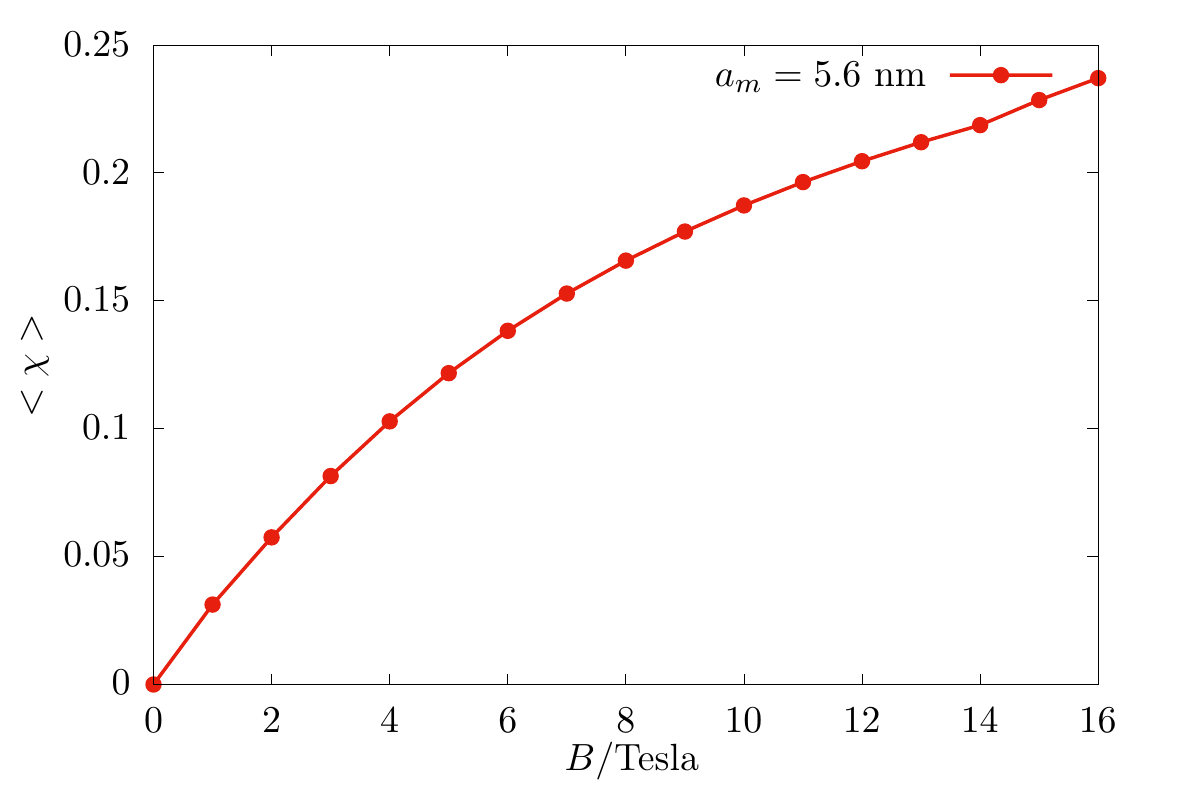}
\caption{ The chiral order $<\chi >$ at $a_{m} = 5.6$ nm for various $B$.}
\label{Figs_chiral}
\end{figure}

The chiral order is studied on a finite $N=48\times 6$ lattice, which is defined as $< \chi >=1/(N-1)\sum\limits_{\{ ijk \} \in \bigtriangleup / \bigtriangledown  }\mathbf{S}_{i}\cdot (\mathbf{S}_{j}\times \mathbf{S}_{k})$ due to the open boundary. As shown in Fig.~\ref{Figs_chiral}, the chiral order smoothly increases as $B$ increases from the $120^{\circ}$ N\'{e}el phase to the CSL phase and it does not show a critical behavior near the phase transition point of around $B=5$ T. We believe larger systems may be needed to see the critical behavior. The phase boundary between the $120^{\circ}$ N\'{e}el phase and the CSL phase is determined by the characteristic counting in the entanglement spectrum.

\end{document}